\documentclass[fleqn,usenatbib,useAMS]{mnras}

\usepackage{graphicx}	
\usepackage{amsmath}	
\usepackage{amssymb}	
\usepackage{multicol}        
\usepackage{bm}		
\usepackage{pdflscape}	
\usepackage{longtable}

\usepackage{hyperref}

\usepackage{natbib}
\setcitestyle{authoryear,open={(},close={)}}

\def\Tend{150\,\mathrm{Myr}}
\def\Mbh{M\mathrm{_{BH}}}
\def\NdotTDE{\dot{N}_{\mathrm{TDE}}}
\def\Mdotbh{\dot{M}_{\mathrm{BH}}}
\newcommand{\NTDE}{N_{\mathrm{TDE}}}
\newcommand{\Rinf}{R_{\mathrm{inf}}}
\newcommand{\Rt}{R_{\mathrm{t}}}
\newcommand{\Rp}{R_{\mathrm{p}}}
\newcommand{\Rh}{R_{\mathrm{h}}}
\newcommand{\rc}{r_{\mathrm{c}}}
\newcommand{\trel}{t\mathrm{_{rel}}}

\newcommand{\fcTCE}{f_\mathrm{c}^\mathrm{TCE}}

\newcommand{\Nb}{N_{\mathrm{b}}}

\usepackage[T1]{fontenc}
\usepackage{ae,aecompl}

\usepackage{newtxtext,newtxmath}

\title[Growth of massive black holes in dense star clusters]{The growth of intermediate mass black holes through tidal captures and tidal disruption events}
\author[F. P. Rizzuto]{Francesco Paolo Rizzuto$^{1,2}$\thanks{Contact e-mail: \href{mailto:}{francesco.rizzuto@helsinki.fi}},
Thorsten Naab$^{2}$, Antti Rantala$^{2}$,  Peter H. Johansson$^{1}$,  \newauthor
Jeremiah P. Ostriker$^{3, 4}$, Nicholas C. Stone$^{5}$, Shihong Liao$^{1}$, Dimitrios Irodotou$^{1}$
\\ 
$^{1}$Department of Physics, University of Helsinki, Gustaf Hällströmin katu 2, FI-00014, Helsinki, Finland
\\
$^{2}$Max-Planck Institute for Astrophysics, Karl-Schwarzschild-Str. 1,D-85741 , Garching, Germany
\\
$^{3}$Department of Astronomy, Columbia University, New York, NY 10027, USA
\\
$^{4}$Department of Astrophysical Sciences, Princeton University, Princeton, NJ 08544, USA
\\
$^{5}$Racah Institute of Physics, The Hebrew University, Jerusalem, 91904, Israel
}

\date{\today}

\pubyear{2020}

\begin{document}
\label{firstpage}
\pagerange{\pageref{firstpage}--\pageref{lastpage}}
\maketitle

\begin{abstract}
We present $N\mathrm{-body} $ simulations, including post-Newtonian dynamics, of dense clusters of low-mass stars harbouring central black holes (BHs) with initial masses of 50, 300, and 2000 $\mathrm{M_{\odot}}$. The models are evolved with the $N\mathrm{-body} $ code
\textsc{bifrost} to investigate the possible formation and growth of massive BHs by the tidal capture of stars and tidal disruption events (TDEs). We model star-BH tidal interactions using a velocity-dependent drag force, which causes orbital energy and angular momentum loss near the BH. 
About $\sim 20-30$ per cent of the stars within the spheres of influence of the black holes form Bahcall-Wolf cusps and prevent the systems from core collapse. Within the first 40 Myr of evolution, the systems experience 500 – 1300 TDEs, depending on the initial cluster structure. Most ($> 95$ per cent) of the TDEs originate from stars in the Bahcall-Wolf cusp. We derive an analytical formula for the TDE rate as a function of the central BH mass, density and velocity dispersion of the clusters ($\NdotTDE\propto \Mbh \rho \sigma^{-3}$). We find that TDEs can lead a 300 $\mathrm{M_{\odot}}$ BH to reach $\sim 7000 \mathrm{M_{\odot}}$ within a Gyr. This indicates that TDEs can drive the formation and growth of massive BHs in sufficiently dense environments, which might be present in the central regions of nuclear star clusters.

\end{abstract}

\begin{figure*}
	\includegraphics[width=0.98\textwidth]{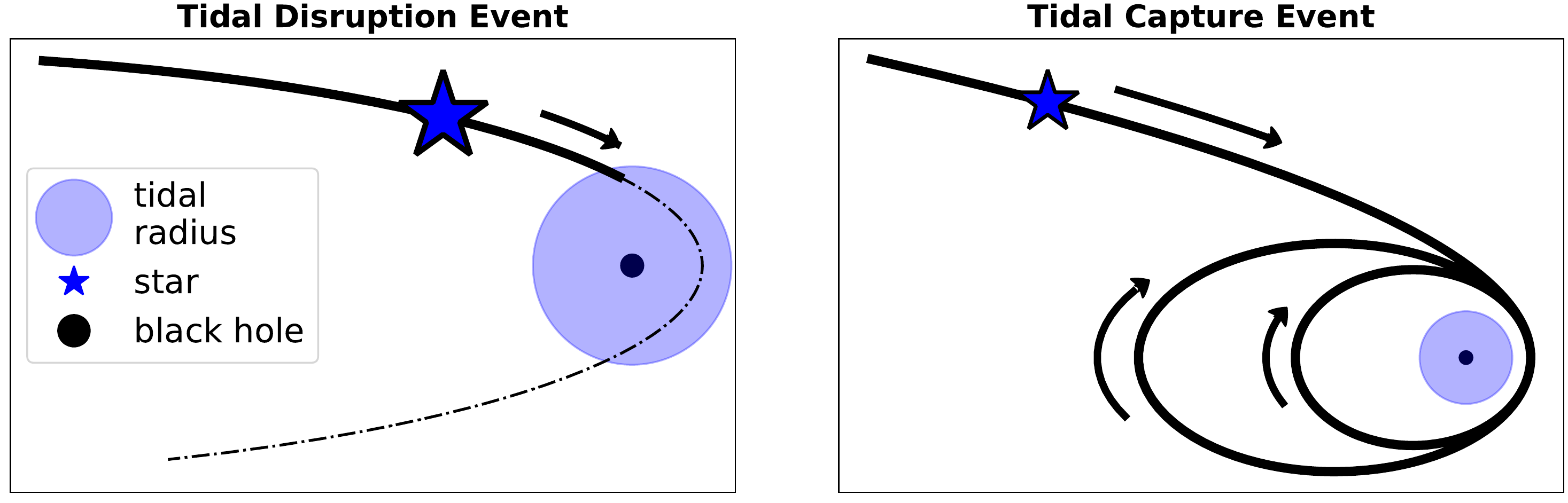}
\caption{\textit{Left panel}: schematic illustration of a TDE. 
A star approaches a black hole moving in a wide orbit with a pericentre smaller than the tidal radius $\Rt$ 
(identified with the blue sphere in the plot). As soon as the star - BH separation is smaller than $\Rt$ the star experiences a tidal pull strong enough to tear it apart.
\textit{Right panel}: schematic illustration of a TCE. In this case, the star follows an unbound orbit with pericentre $\gtrsim \Rt$. Therefore the star is not destroyed. However, the BH is sufficiently close to cause internal oscillations in the star, converting a fraction of the initial orbital energy of the star into internal energy. The orbit becomes bound: the star is "captured". If the same process repeats at every pericentre passage, the stellar orbit will shrink over time, and can eventually lead to most of the star's mass accreting onto the BH.}\label{fig:1_sketch_TDE_TCE}
\end{figure*}
\section{Introduction}
It is well established that most massive galaxies host in their centre black holes with masses above $10^6 \mathrm{M_{\odot}}$, also known as supermassive black holes (SMBHs) \citep{Magorrian1998, Kormendy2013}. Historically, the first strong indications of their presence were active galactic nuclei (AGNs), for which the high luminosity and electromagnetic spectrum were explainable only by the accretion of matter by supermassive compact objects \citep[][]{Lynden-Bell1969}.
In recent years, observers utilising the Event Horizon Telescope have even been able to produce the first direct image of the shadow of the $\approx6 \times 10^9 \mathrm{M_{\odot}}$ SMBH  at the centre of the Messier 87 galaxy \citep{EHT2019} as well as Sagittarius A*, the $\approx 4\times10^6\mathrm{M_{\odot}}$ SMBH in the centre of the Milky Way \citep{EHT2022}.
Similarly, the existence of stellar mass BHs $\Mbh<100\mathrm{M_{\odot}}$, the end product of massive star evolution, has been confirmed by several X-ray and optical observations \citep{Casares2014, ElBadry2022, Remillard2006} and more recently even by gravitational wave (GW) detections \citep{Abbott2016, Abbott2017, Abbott2021}.

On the other hand, black holes with masses between $10^2 \mathrm{M_{\odot}} \lesssim \Mbh \lesssim 10^6 \mathrm{M_{\odot}}$, referred to as intermediate-mass BHs (IMBHs), remain more elusive.
Over the last decades,  several observations seem to indicate the presence of IMBHs in galactic nuclei. More than thirty years ago, \citet{Kunth1987}
pointed out an AGN in the dwarf galaxy POX 52, which might be powered by a $\sim 10^5 \mathrm{M_{\odot}}$ BH \citep{Barth2004}. 
Similarly, a $10^4-10^5 \mathrm{M_{\odot}}$ BH could be responsible for the
Seyfert activity detected in the nuclear star cluster of the dwarf spiral galaxy NGC 4395 \citep{Filippenko1989, Shih2003, Peterson2005, denBrok2015}                   )
l.77. Many objects with similar masses have been found in several subsequent surveys \citep[see][and references therein]{Greene2020, Reines2022}. Recently, \citet{Gultekin2022} showed that at least five sources located in low-mass galaxies (with stellar masses $<3\times 10^9 \mathrm{M_{\odot}}$), are consistent with active BHs with masses between $10^{4.9}\mathrm{M_{\odot}}<\Mbh<10^{6.1}\mathrm{M_{\odot}}$. The presence of such BHs might not be limited to the centres of galactic nuclei; a few of the most massive globular clusters might also be hosting objects of similar masses \citep{Farrell2014, Pechetti2022}. Recently, a tidal disruption event (TDE) was discovered \citep{Lin2018} in a dense stellar object of mass $\sim 10^7 \mathrm{M_{\odot}}$, likely either a large globular cluster or a tidally stripped satellite nucleus.  X-ray continuum fitting indicates that the TDE was powered by an IMBH of mass $\sim 10^4 \mathrm{M_{\odot}}$ \citep{Wen2021}. 

The physical processes leading to the formation of these BHs are still debated in the literature \citep[see][for a comprehensive review]{Volonteri2021}. One of the most promising scenarios for growing such massive BHs is the collisional scenario. Low-mass BHs  (a few tens of solar masses) located in dense stellar environments could grow in mass through stellar and compact object mergers  \citep[see, for example][]{Stone2017}. Gravitational wave observations could have already revealed the tip of the iceberg of this process. The LIGO-Virgo-KAGRA interferometers have thus far detected about  ninety BH coalescences, a few of which generated BHs  with masses above  $\gtrsim 100 \mathrm{M_{\odot}}$  \citep{Abbott2021}.  Being the most massive members of many star clusters, stellar-mass BHs are expected to sink in the very centre of the host stellar system. There, if the density is high enough, a chain of hierarchical mergers could produce BHs with masses around $10^2 \mathrm{M_{\odot}} - 10^4 \mathrm{M_{\odot}}$  \citep{Atallah2022, Arca-Sedda2021, Fragione2022, Mapelli2021, Rizzuto2021} or even $10^5 \mathrm{M_{\odot}}$  according to \citet{Antonini2019}. The main obstacle to this hierarchical formation path is recoil caused by the anisotropic emission of gravitational radiation. The final product of compact object mergers is expected to receive GW-induced velocity kicks that range from a few tens up to $5000$ km/s depending on the mass ratio and the spins of the merging bodies \citep{Campanelli2007, Lousto2010, Lousto2019}. Systems with low escape velocities, such as globular clusters, are unlikely to retain the final product of repeated BH mergers \citep{Arca-Sedda2021, Gerosa2019}. Only dense stellar environments with an escape velocity in excess of $v_{\mathrm{esc}} \gtrsim 100$ km/s have a non-negligible chance to grow IMBHs through hierarchical collisions \citep{Gerosa2019, Mapelli2021}. Nuclear star clusters are the only class of stellar systems that fulfill this condition \citep{Neumayer2020}. 

In star cluster environments, stellar BHs coexist with stars, which provide an alternative merger channel for BH mass growth, especially in young compact clusters which are populated by many massive stars. Direct $N\mathrm{-body} $ simulations presented in \citet{Zwart2004} showed that young dense low-metallicity star clusters, if compact enough, can trigger repeated stellar mergers and generate very massive stars (VMSs) with masses up to $\sim 1000 \mathrm{M_{\odot}}$,
which collapse directly into massive BH without losing mass in the process.
Recent direct $N\mathrm{-body} $ simulations evolve similar systems using up-to-date stellar evolution recipes and confirm the collisional formation scenario of VMSs \citep{DiCarlo2021, Rizzuto2022}. However, they also indicate that such a process can lead at most to BHs with masses of a few $100 \mathrm{M_{\odot}}$  in agreement with the latest Monte-Carlo simulations \citep{Kremer2020, Gonzalez2022}. The evolution of VMSs is highly uncertain. \citet{Glebbeek2009} argues that the final product of massive stellar mergers is affected by enhanced mass loss that prevents VMSs from generating BH more massive than $10^2 \mathrm{M_{\odot}}$. 

Young massive star clusters may have another channel for massive BHs to grow: mergers between BHs and massive stars.
Such events have been reported in both direct $N\mathrm{-body} $ simulations \citep{Mapelli2016, Rizzuto2021, Rizzuto2022} and Monte-Carlo simulations \citep{Giersz2015}. 
In \citet{Rizzuto2021} we show that BHs up to  $\sim 3\times 10^2 \mathrm{M_{\odot}}$ can form even when the VMS direct collapse channel is  suppressed\footnote{With the stellar evolution recipes adopted in \citet{Rizzuto2021} VMSs above  $>10^2 \mathrm{M_{\odot}}$ form BHs lighter than $30 \mathrm{M_{\odot}}$.}  as long as stellar BHs can accrete a significant fraction of the stellar material when colliding with a massive star. 

Young massive star clusters can generate massive BHs only during the early stage of their evolution when they are still dense, and massive stars are still present in the system. As soon as stars undergo supernova explosions, the clusters lose a large fraction of their central mass and subsequently experience a rapid expansion. The most massive stars terminate their life, collapsing into stellar black holes.
As these BHs are the heaviest objects left in the system, they segregate in the cluster core and form a compact subsystem.
It is well established in the literature that the BH subsystem prevents the cluster core-collapse, drives stars away from the centre, and can even remove them from the cluster\citep{Merritt2004, Breen2013}.
At the same time, also BHs can be dynamically ejected during repeated few-body close encounters. Because of these interactions, over time, star clusters can lose most of their BHs.
Which of the two components, BHs or stars, will dissolve first depends on whether the cluster is tidally filling\footnote{A cluster is tidally filling when it contains stars all the way to its tidal radius, otherwise it is tidally underfilling. Typically compact and dense systems are tidally underfilling while more dilute clusters with a moderate initial central density are tidally filling.} at the beginning or during its evolution \citep{Giersz2019}.
In tidally filling clusters, stars escape faster than  BHs.  These systems evolve toward a state in which the BHs constitute the majority of their mass. 
The low-density globular cluster Palomar 5 has most likely entered this phase since direct $N\mathrm{-body} $ simulations suggest that about $20$ per cent of its mass is in the form of BHs \citep{Gieles2021}.  

On the other hand, in tidally underfilling clusters, the BH component evaporates first, leaving behind only a few black holes immersed in a sea of low-mass stars. An example of such a system is the $\sim 10^5 \mathrm{M_{\odot}}$ globular cluster NGC 6397,       
which hosts no more than a dozen of BHs in its core \citep{Weatherford2020}.
The presence of an IMBH would further catalyse the removal of the BH subcluster, as simulations show that the massive compact object is likely to rapidly expel most BHs and BH binaries during few-body interactions \citep{Giersz2015}.
In this case the star cluster would quickly evolve into a system consisting of low-mass stars enclosing a central massive BH. During this stage the BH can slowly consume the surrounding stars through tidal disruption events or tidal capture events (TCEs). The former occurs when the BH is close enough to a star to exert a tidal pull that rips it apart (left panel of Fig. \ref{fig:1_sketch_TDE_TCE}). 
The latter case occurs when an unbound star is sufficiently near to a BH to experience a strong enough tidal perturbation to become bound.
The tidal force deforms the nearby star, so part of its initial orbital energy is transferred to its internal energy. This mechanism can convert unbound stars onto a bound orbit. When this happens, the star is said to be "captured" by the BH (see right panel of Fig. \ref{fig:1_sketch_TDE_TCE} as an illustration of a TCE). After being captured, the star can feed the compact object in various ways, such as, partial disruption events at each pericentre passage \citep{Kremer2022}, mass transfer events, or even dynamically induced mergers. 
TCEs were highlighted for the first time in \citet{Fabian1975}; 
 \citet{Press1977} provided the first mathematical description of these events, later on, refined by \citet{Lee1986}.  
 
Since the tidal capture mechanism has a larger cross-section than the tidal disruption process, it is likely to impact BH mass growth in dense stellar environments significantly. The analytical work led by \citet{Stone2017} provided, for the first time, a comprehensive study on the critical role played by TCEs and TDEs in forming massive BHs in galactic nuclei.
 Their work showed that a low mass BH located in the centre of a dense stellar environment can reach, within a Hubble time, up to $10^6 \mathrm{M_{\odot}}$ through tidal disruptions and tidal capture runaway collisions. According to their calculation, the rapid growth can be triggered only in clusters with a core density of $n_{\mathrm{c}}>10^7/$pc$^3$ and a core velocity dispersion of $\sigma_{\mathrm{c}} > 40$ km/s.  
 
In this work, we explore the tidal collisional runaway scenario using direct $N\mathrm{-body} $ simulations. Therefore, we evolve star clusters initialised according to the density and velocity dispersion criteria given in \citet{Stone2017}, and study the dynamic mechanisms responsible for the growth of massive BH.
In Section \ref{sec:Mathods}, we describe in detail the numerical integrator and the  tidal interactions prescriptions adopted for our investigation. In Section \ref{sec:InitialConditions}, we discuss the intial conditions of our clusters. In Section \ref{sec:Results} we present the results of our simulations and in Section \ref{sec:Conclusions} we offer a summary of our main results.

\section{Methods}  \label{sec:Mathods}

In order to investigate the growth of massive BHs in compact stellar environments through TDEs and TCEs, we run five direct $N\mathrm{-body} $ simulations of dense star clusters consisting of a central BH surrounded by low-mass stars.  We evolve the systems for at least $40$ Myr using the direct $N\mathrm{-body} $ integrator \textsc{bifrost} \citep{Rantala2022}, which we describe in detail in the following Subsection. 
We include in \textsc{bifrost} prescriptions for modelling the effect of tidal interactions during close encounters using a drag force as described in Subsection \ref{subsec:drag force}. 
\subsection{The \textit{N}-body code}
In this study we used the novel GPU-accelerated direct-summation $N\mathrm{-body} $ simulation code \textsc{bifrost} \citep{Rantala2022} which is based on the earlier FROST code \citep{Rantala2021}. The code uses a hierarchical implementation (HHS-FSI, \citealt{Rantala2021}) of the fourth-order forward symplectic integrator (FSI, e.g. \citealt{Chin1997,Chin2005,Dehnen2017}). 
In addition to the Newtonian accelerations, FSI uses additional so-called gradient accelerations to cancel second-order error terms with strictly positive integrator sub-steps. This is different to the widely used Yoshida-type symplectic integrators \citep{Yoshida1990} which always contain negative sub-steps in integrator orders higher than two. For the Kepler problem it has been shown that fourth-order symplectic integrators with strictly positive sub-steps outperform the integrators that include negative sub-steps \citep{Chin2007}. Compared to the common block time-step scheme widely used in $N\mathrm{-body} $ simulations \citep{Aarseth2003}, HHS-FSI is manifestly momentum-conserving due to pair-wise acceleration calculations and synchronized kick operations. In addition, rapidly evolving parts of the simulated systems decouple from slowly evolving parts in the hierarchical integration, making the approach especially efficient for $N\mathrm{-body} $ systems with a large dynamical range \citep{Pelupessy2012}. \\
Besides the forward integrator, \textsc{bifrost} includes regularised \citep{Rantala2020} and secular integration techniques for subsystems, i.e. binaries, triple systems and small clusters around massive black holes. The equations of motion of the particles in the subsystems are post-Newtonian (PN) up to order PN3.5 using the formulas of \citet{Thorne1985,Blanchet2006}. For binary systems the post-Newtonian equations of motion enable relativistic precession effects due to conservative even PN terms, as well as orbit circularisation and inspiral due to radiation-reaction forces due to gravitational wave emission. \textsc{bifrost} can evolve massive star clusters with high numerical accuracy containing up to a few million stars with an arbitrary fraction of primordial binaries \citep{Rantala2022}, a feature which only few current codes share \citep{Wang2020}. \\
The basic \textsc{bifrost} version has four prescriptions for mergers. First, two particles are merged if their gravitational-wave inspiral timescale becomes shorter than their current time-steps. We also merge particles if their mutual separation is shorter than the relativistic innermost stable circular orbit. \textsc{bifrost} also includes a simple prescription for tidal disruption events which is significantly extended for this study as described in the next sections. Finally, two stars are merged if they directly collide, i.e. their radii overlap. Table \ref{table:bifrost_parameters} reports the main \textsc{bifrost} code parameters used in our runs. 
\begin{table}
\begin{tabular}{l l l}
\hline
\textsc{bifrost} user-given parameter & symbol & value\\
\hline
forward integrator time-step factor & $\eta_\mathrm{ff}$, $\eta_\mathrm{fb}$, $\eta_\mathrm{\nabla}$ & $0.2$\\
subsystem neighbour radius & $r_\mathrm{rgb}$ & $0.008$ pc\\
regularization GBS tolerance  & $\eta_\mathrm{GBS}$ & $10^{-10}$\\
regularization GBS end-time tolerance & $\eta_\mathrm{endtime}$ & $10^{-4}$\\
regularization highest PN order &  & PN3.5\\
secular integration threshold & $N_\mathrm{orb,sec}$  & $10$\\
secular highest PN order &  & PN2.5\\
\hline
\end{tabular}
\caption{The main user-given code \textsc{bifrost} parameters used in the simulation runs in this study. The parameter definitions correspond to the ones in \citet{Rantala2022}.}
\label{table:bifrost_parameters}
\end{table}
\begin{figure*}
	\includegraphics[width=\textwidth]{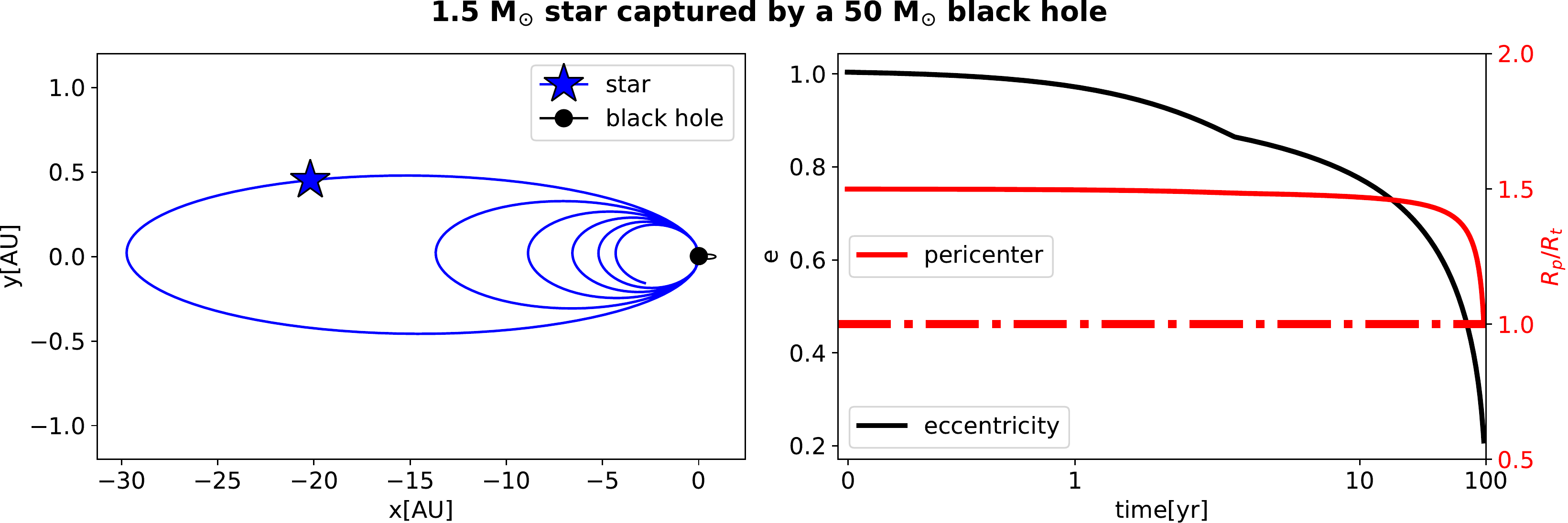}
	\caption{{\it Left panel}: Evolution of an initially unbound orbit of a tidally affected $1.5 \mathrm{M_{\odot}}$ star around a $50 \mathrm{M_{\odot}}$ black hole. Due to tidal energy losses (implemented in the simulation as a drag force), the star loses orbital energy and its orbit circularises. {\it Right panel}: Time evolution (in log-scale) of the orbital eccentricity (black) and the pericentre distance (solid red line) of the star. In the final stages the orbit circularises and the pericentre distance shrinks. Once the pericentre becomes smaller than the tidal radius (red dot-dashed line) the star is disrupted in a TDE.}  
	\label{fig:2_TCE_15_star}
\end{figure*}
\subsection{Prescription for tidal disruption and mass accretion}
In \textsc{bifrost}, a BH destroys a star when $r$, the distance between the two interacting objects is smaller than the tidal radius $\Rt$.
To compute $\Rt$ we adopt the criterion given in \citet{Kochanek1992}:
\begin{equation}
\Rt=1.3R_*\left(\frac{\Mbh + m_*}{2m_*}\right)^{1/3},
\end{equation}
where $R_*$ and $m_*$ are the radius and the mass of the destroyed star, respectively, while $\Mbh$ is the BH mass. 
To ensure the codes does not miss any TDEs, we decrease the time steps of each interaction with pericentre $\Rp< 3 \Rt$, for the code to check the condition $r < \Rt$ exactly at pericenter passage. With this restriction, the regularised integrator is less efficient, but fortunately, these close interactions are infrequent. Therefore they have a negligible impact on the overall performance of the code.

After a TDE, a fraction of the stellar material is expected to be ejected at high velocity, while the remaining mass remains bound to the BH and eventually forms an accretion disk around it. Recent hydrodynamic simulations of tidal disruption events between stellar BHs and main-sequence stars show that the fraction of stellar material bound to the BH could be close to unity \citep{Kremer2022}. However, these simulations do not model the accretion phase in which  
a significant fraction of the initially bound gas may get unbound, especially if the accretion occurs at a super-Eddington rate \citep{Ayal2000, Bonnerot2020,  Dai2018, Metzger2016, Steinberg2022, Toyouchi2021}.
In general, many aspects of the accretion phase are poorly constrained. For example, it is not very well understood  
how much stellar mass ends up into the BH and how much is lost, and the rate at which the accretion should occur is still debated in the literature.
In this work we follow the simple estimate
in \citet{Rees1988} and assume that in a tidal disruption event, $50$ per cent of the stellar mass is accreted by the BH instantly\footnote{During these instantaneous accretion events we assume that linear momentum is conserved.}, and the other $50$ per cent is instantly removed from the cluster. 
We discuss how this simplification affects our results in Subsection \ref{sub:mass_growth}. 

\subsection{Tidal interaction energy loss}
When a star of mass $m_*$ moves in an orbit with a pericentre slightly larger than the tidal radius $\Rp \gtrsim \Rt$, tidal forces are not strong enough to rip the star apart, but they can still deform the star triggering internal oscillations. 
Consequently, a fraction of the orbital energy of the star is lost in the internal energy of the star. In other words, tidal interactions force stars to deviate from their original orbits: bound eccentric orbits become more bound and circular, and unbound parabolic or hyperbolic orbits become bound.

To estimate the energy lost during a parabolic encounter, \citet{Press1977} approximated the oscillations amplitude of the perturbed star through an expansion in spherical harmonics. With their procedure, the fraction of orbital energy deposited into the star is:
\begin{equation}\label{eq_press_teukolsky_1977}
\Delta E = \int_{-\infty}^{+\infty}\frac{\mathrm{d}E}{\mathrm{d}t}\mathrm{d}t \approx \frac{Gm_*^2}{R_*}\left(\frac{\Mbh}{m_*} \right)^2 \sum_{l=2}^{\infty} \left(\frac{R_*}{\Rp} \right)^{2l+2}T_l(\eta).
\end{equation}
Here $T_l$ are dimensionless coefficients that depend on the internal structure of the deformed object. They are functions of the parameter $\eta$ defined as:
\begin{equation}
\eta \coloneq \left(\frac{m_*}{m_*+\Mbh} \right)^{1/2}\left(\frac{\Rp}{R_*}\right)^{3/2}.
\end{equation}
As follows from its definition, $\eta$ indicates the duration of the pericentre passage with respect to the hydro-dynamical timescale of the star $m_*$.
In practical applications, only $l=2$ and $l=3$ are taken into account. $T_0=0$ and higher terms give a negligible contribution to the final value of $\Delta E$. The values of $T_2$ and $T_3$ have been calculated explicitly for objects with polytropic indices $n=1.5, 2$ and $n=3$ \citep[see][]{LeeandOstriker1986, Ray1987}. Based on these values, \citet{PortegiesZwart1993} provide
fitting functions to rapidly estimate $T_2$ and $T_3$ during close encounters in $N\mathrm{-body} $ simulations. 

The formula shown in Eq. \ref{eq_press_teukolsky_1977}, has been derived exclusively for parabolic encounters ($e=1$).
To extend this formulation for eccentricities larger or smaller than $1$, we utilise the prescription given in appendix A of \citet{Mardling2001}, which provides a generalisation of Eq. \ref{eq_press_teukolsky_1977}
by introducing a new expression for $\eta$ (indicated with $\zeta$) that depends explicitly on the eccentricity of the orbit:
\begin{equation}\label{eq:zeta}
\zeta \coloneq \eta \left( \frac{2}{1+e} \right)^{\alpha(\eta)/2}
\end{equation}
where $\alpha(\eta)=1 + \frac{1}{2}\left|\frac{\eta -2}{2}\right|$, therefore for $e=1$ we restore the original formulation. Numerical tests show that using $T_l(\zeta)$ instead of $T_l(\eta)$, the tidal evolution of hyperbolic orbits ($e\gtrsim1$) is consistent with a more accurate model for tidal interactions presented in \citet{Mardling1995}. In addition, $T_l(\zeta)$ should give more reliable results for bound orbits with $e\gtrsim0.5$ 
\citep[see][]{Mardling2001}. 
To estimate the tidal energy loss during a  BH - star close encounter in our simulations we first compute $\zeta$ using Eq.
\ref{eq:zeta} and then evaluate $T_2(\zeta)$ and $T_3(\zeta)$ using the interpolation formulas given in \citet{PortegiesZwart1993}.  
Finally, we use Eq. \ref{eq_press_teukolsky_1977} to estimate the energy loss in tidal interactions.
For orbits with $e<0.5$, we assume the tides switch from the dynamical to the equilibrium regime. The amount of tidal energy dissipation, therefore, decreases with decreasing eccentricity. To take into account this effect, we multiply 
the energy dissipation $\Delta E$ by the eccentricity $e$ ($\Delta E=e\times \Delta E$) similarly to what was done by \citet{Baumgardt2006}.

\subsection{Modelling tidal interactions with a drag force}
\label{subsec:drag force}

In the previous subsection, we described the procedures adopted to estimate $\Delta E$, the amount of energy dissipated during tidal interactions. There remains the problem of how to use  $\Delta E$ to 
change the orbit of the two closely interacting objects. 
The standard approach assumes that all the energy $\Delta E$ is instantly emitted at each pericentre passage. Then under the assumption that the angular momentum is conserved, one can derive the new eccentricity and semi-major axes from the present one \citep{Mardling2001} and updates the orbit accordingly. The entire procedure must be implemented outside the regularised integrator. In practice, to resolve the trajectory, the integrator must evolve the orbit to the pericentre, apply the orbital change due to tidal energy loss, then continue the evolution to the next pericentre passage.
This procedure is relatively straightforward to implement in isolated two-body encounters, but it becomes very cumbersome when three or more objects are involved in the interaction.  
Integrating the effects of tidal interactions directly in the regularisation would be significantly more convenient. 
For this reason, we decided to model tidal interactions using a drag force following  \citet{Samsing2018}: 
\begin{equation}\label{eq:dragforce4}
	\vec{F}_{\text{t}}(r,v) = -C \frac{\vec{v}}{r^4}
\end{equation}
where $r$ and $v$ are the relative separation and relative velocity of the two tidally interacting bodies, respectively, while $C$ is a normalisation factor: 
\begin{equation}
	\int_{\mathrm{orb}} F_t(r,v) dr = \Delta E.
\end{equation}
Here $\Delta E$ is the same quantity computed in the instant emission treatment (Eq. \ref{eq_press_teukolsky_1977}). With this definition, $C$ ensures the two method dissipate the same amount of energy per orbit. 

The force expression of Eq. \ref{eq:dragforce4} closely resembles the formula of the dissipative PN 2.5 radiation-reaction force.
For this reason, Eq. \ref{eq:dragforce4}  can be included with little effort in the \textsc{bifrost} regularised integrator alongside the PN terms.
The strong dependence of $F_{\text{t}}$ on $r$, the distance of the two interacting bodies, ensures that tidal interactions are effectively activated during very close encounters and are negligible at large distances.

\begin{figure*}
	\includegraphics[width=\textwidth]{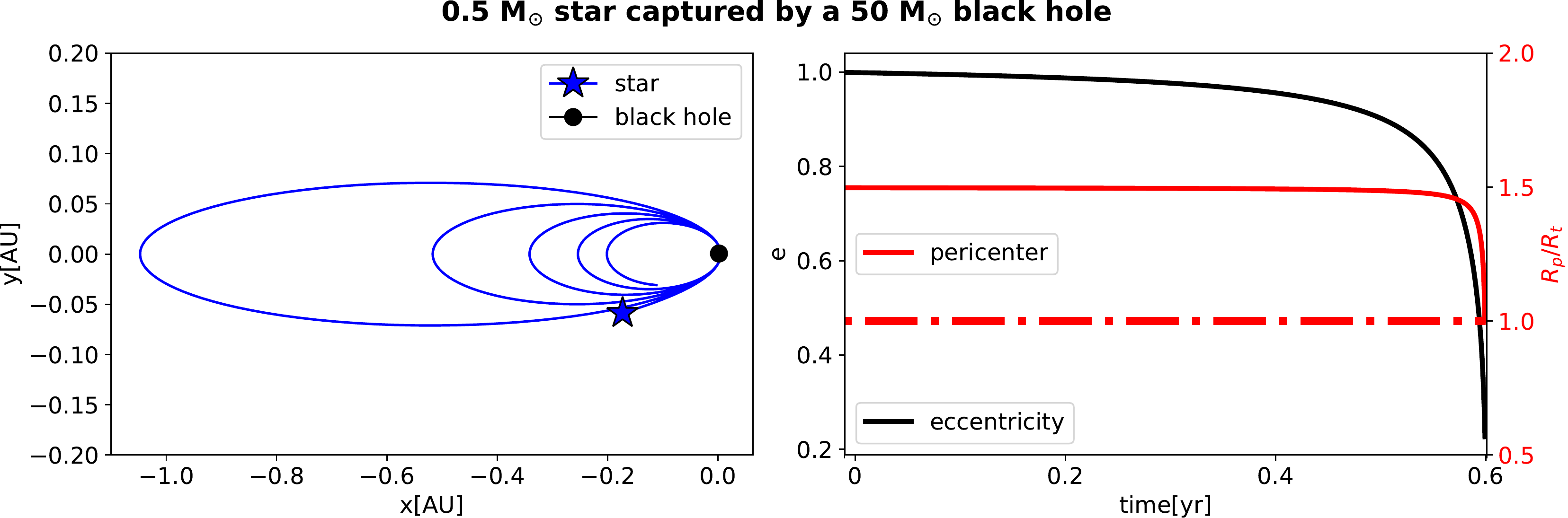}
	\caption{{\it Left panel}: Evolution of an initially unbound orbit of a tidally affected $0.5 \mathrm{M_{\odot}} $ star around a $50 \mathrm{M_{\odot}} $ black hole. The evolution is similar to the $1.5 \mathrm{M_{\odot}} $ star but on shorter timescales. {\it Right panel}: Time evolution (in linear-scale) of the orbital eccentricity (black) and the pericentre distance (solid red line) of the star. In the final stages the orbit circularises and the pericentre distance shrinks. Once the pericentre becomes smaller than the tidal radius (red dot-dashed line) the star is disrupted in a TDE.}
	\label{fig:3_TCE_05_star}
\end{figure*}
\subsection{Tidal interactions drag force and orbital evolution}
\label{sub:orbit under drag force}
The instant emission prescription for tidal interactions assumes conservation of angular momentum; therefore the pericentre increases until the orbit circularizes \citep[see][for more details]{Mardling2001}.
On the contrary, the tidal interaction drag force does not conserve angular momentum. Therefore, with this method the pericentre of tidally captured stars decreases over time until $\Rp < \Rt$. In other words, in the absence of external perturbers, the tidal interaction drag force drives every TCE into a TDE . The time required to bring a captured star to tidal disruption depends on the star's initial pericentre, initial eccentricity and internal structure. In this work, we model low-mass stars ($m_*<0.7 \mathrm{M_{\odot}}$) with the polytropic index $n=1.5$ and massive stars ($m_*> 0.7 \mathrm{M_{\odot}}$) with polytropic index $n=3.0$. Consequently, massive stars are more compact and less affected by tidal perturbations than low-mass stars. Fig. \ref{fig:2_TCE_15_star} shows the orbital evolution (left panel) of a $1.5 \mathrm{M_{\odot}}$ star captured by a $50 \mathrm{M_{\odot}}$ BH. The star initially moves in a quasi-parabolic orbit with $e=1.0001$ and $\Rp=1.5 \Rt$. In the beginning, the eccentricity decreases while the pericentre stays constant. In the last stage of the evolution, $\Rp$ drops rapidly and reaches the tidal radius. It takes about $100$ yr for the star to be destroyed (see right panel of Fig. \ref{fig:2_TCE_15_star}). The $0.5 \mathrm{M_{\odot}}$ star shown in Fig. \ref{fig:3_TCE_05_star} experiences a similar evolution. However, the effect of the drag force is much stronger when acting on the $0.5 \mathrm{M_{\odot}}$ star. Consequently, the low-mass star spiral towards the BH in just 0.6 years.

As shown in Figs. \ref{fig:2_TCE_15_star} and \ref{fig:3_TCE_05_star}, the stars, after being captured, undergo a intial phase of circularisation (the eccentricity drops) while the pericentre remains unchanged. Only in the last phase of evolution, also the pericentre begins to decrease. 
We will use this feature of the drag force to identify all the TCEs that occurred in our simulations (see Subsection \ref{sub:direct TDE}).

To summarise, in our simulations, we model the effect of tidal interactions using the drag force described by Eq. \ref{eq:dragforce4}. With such a prescription, every TCE will produce a TDE in less than $\lesssim 10^3$ yr.
It must be said that the outcome following a TC event is highly debatable. In this paper, we assume that every tidal capture leads rapidly to tidal disruption. However, this may not be true. A star, after being captured, if not affected by external perturbations, follows a bound but very eccentric orbit.
If internal mechanisms (such as viscosity) of the star are adequately efficient to rapidly dissipate the oscillatory motion of the star, at the second pericentre passage, the star will experience again a strong tidal interaction that leads to orbital energy loss and consequently to the shrinking of the semi-major axis. The process repeats until the star gets destroyed and swallowed by the BH, or the two bodies circularise. If the spin-orbit coupling is inefficient, orbital angular momentum is conserved, and the star circularises around the BH. If the spin-orbit interaction is efficient, a fraction of orbital angular momentum can be transferred efficiently into the stellar spin.  

Another thing to consider is the possible expansion of the stars after the pericentre passage due to the gain in internal energy. In this scenario,  the radius of the star $R_*$ will increase, resulting in an increased tidal radius: the star will be tidally disrupted after a few pericentre passages.
Another possible scenario is described in \citet{Mardling2001}, where they argue that if the oscillations induced by tidal interaction are not damped efficiently, TCEs might lead to chaotic evolution. In fact, the pericentre passages that follow the capture can further excite but also damp oscillation modes in the deformed star. In other words, the orbit and the star can randomly exchange energy packages in both directions. The resulting trajectories of the object are unpredictable and resemble a random walk. 
Recent hydrodynamical simulations indicate that captured stars that experience partial tidal disruptions to be completely destroyed after a few pericentre passages \citep{Kremer2022}. Unfortunately, the simulations sample is too small to constrain the fate of tidally captured stars extensively.

\section{Initial Conditions}
\label{sec:InitialConditions}

 \begin{table}

\begin{tabular}{cccccc}
\hline
Name &  $\Mbh$ & $\Rh$ &   $\rc$ &      $n_{\mathrm{c}}$ &  $\sigma_{\mathrm{c}}$ \\
-    & $\mathrm{M_{\odot}}$  & pc   & pc        & $\#$ stars / pc $^3$       & km/s   \\
\hline
R04M300  &  300 & 0.4 & 5.4e-03 & 1.4e+09 & 42.3 \\
R06M300  &  300 & 0.6 & 7.5e-03 & 1.7e+08 & 32.9 \\
R08M300  &  300 & 0.8 & 1.1e-02 & 6.8e+07 & 28.5 \\
R06M50   &   50 & 0.6 & 1.7e-02 & 5.5e+07 & 26.3 \\
R06M2000 & 2000 & 0.6 & 3.3e-03 & 1.7e+09 & 90.4 \\
\hline
\end{tabular}
\caption{Initial clusters and BH properties: Name: name of the model; $\Mbh$: initial BH mass; $\Rh$: initial half mass radius;  $r\mathrm{_c}\,$: initial core radius; $n\mathrm{_c}\,$: initial central particle density;  $\sigma_{\mathrm{c}}$: initial velocity dispersion. }
\label{table:initial_conditions}
\end{table}
\begin{table*}
\begin{tabular}{cccccccc}
\hline
Name & t &  $\Mbh$ & N$_{\text{TDE}}$ &   $\Rh$ &     $\rc$ &   $\rho_\mathrm{c}$ &  $\sigma_{\mathrm{c}}$ \\
- & Myr    & $\mathrm{M_{\odot}}$   & - & pc  & pc        & $\mathrm{M_{\odot}}$ / pc $^3$       & km/s   \\
\hline
R04M300  &  41.2 & 1152 & 1239 & 0.5 & 2.0e-02 & 8.8e+06 & 33.3 \\
R06M300  & 148.0 & 1329 & 1487 & 0.9 & 6.6e-02 & 8.5e+05 & 23.6 \\
R08M300  &  41.1 &  686 &  532 & 0.9 & 5.3e-02 & 1.6e+06 & 21.4 \\
R06M50   &  41.9 &  556 &  665 & 0.7 & 2.9e-02 & 5.0e+06 & 25.1 \\
R06M2000 &  41.7 & 2844 & 1342 & 0.8 & 1.6e-02 & 1.3e+07 & 51.3 \\
\hline
\end{tabular}
\caption{Cluster and BHs properties at the end of each simulation: Name: name of the model; t: simulation run time.  $\Mbh$: final BH mass; N$_{\text{TDE}}$: total number of TDE; $\Rh$: final half mass radius; $r\mathrm{_c}\,$: final core radius; $\rho\mathrm{_c}\,$: final central density;  $\sigma$: final velocity dispersion. }
\label{table:final}
\end{table*}

We generated the initial conditions for five very dense star clusters with $256000$ single stars and no primordial binaries using the code \textsc{mcluster} \citep{mcluster}. The masses of the stars are sampled from \citet{Kroupa2001} initial mass function with a range from $0.08 \mathrm{M_{\odot}} $ up to $2.00 \mathrm{M_{\odot}}$\footnote{Having chosen the most massive star to be $2 \mathrm{M_{\odot}}$, allows us to neglect the effect of stellar evolution, and focus only on the dynamical evolution of the systems.}. 
We located at the centre of the cluster single BHs with initial masses equal to $50 \mathrm{M_{\odot}}$, $300 \mathrm{M_{\odot}}$ and $2000 \mathrm{M_{\odot}}$. The five models are initialized with primordial mass segregation and at starting time they follow a \citet{King1966} density profile with W$_0=9$; three of these have an initial half mass radius of $\Rh=0.6$ pc, the other two have a half mass radius respectively equal to $0.4$  pc and $ 0.8$ pc.
From now on, we will refer to the five simulations using the labels reported in Table  \ref{table:initial_conditions}\footnote{The core radius reported in the table is computed using $r\mathrm{_c}=\sqrt{(\sum_i \rho_i^2 r_i^2)/(\sum_i \rho_i^2)}$, where $\rho_i$ is the local density around the $i$-particle and $r_i$ is its distance from the centre.}. 

Thus initialised, the systems have very high core densities (from $\sim 6 \times 10^7$ particles per pc$^{3}$ for R06M50 up to $\sim 2 \times 10^9$ particles per pc$^{3}$ for R06M2000) and high central velocities dispersion (ranging from $30$ km/s up to $90$ km/s) and they might resemble the conditions inside the cores of nuclear star clusters.
The initial conditions we chose for our cluster match the criteria indicated by \citet{Stone2017} to trigger a tidal capture runaway collision. As discussed in the introduction the work by Stone indicate that stellar environments with $n_{\mathrm{c}}> 10^7$ stars per pc$^3$ and a central velocity dispersion $\sigma_{\mathrm{c}}>40 $km/s are expected to trigger tidal capture runaway collisions.

\section{Results}
\label{sec:Results}

All five models maintain a significant tidal disruption rate $\NdotTDE$  throughout the simulation. The final number of TDEs depends strongly on the initial size ($\Rh$) of the cluster.
For instance, R04M300, the most compact model, registers a total of 1239 TDEs in about 40 Myr. On the other hand, in the same evolution time, the least compact model, R08M300 records less than half the TDEs (see Table \ref{table:final}).

The primary mechanism that leads stars being close enough to the BH to be destroyed or captured can be broken down into two parts.
First, dynamical interactions in the vicinity of the BH build up a cloud of bound orbits, which accompany the BH until the end of the simulation.
Once these bound orbits are formed, since they are located in the inner part of the cluster, they are constantly perturbed by many gravitational scattering events. As a consequence of these interactions the bound orbit become over time more bound and very eccentric ($e\sim 1$), resulting in their tidal capture or disruption.
In simple words, the BH first forms rapidly a bound cloud in its vicinity and later feeds on it slowly over time.

We provide a comprehensive description of the main properties of the bound cloud in the different simulations in Subsections \ref{sub:bound_cloud} and \ref{sub:Bahcall Wolf}.
The bound cloud does not only influence the BH mass growth, but it also plays an important role in the evolution of the cluster core
as we discuss in Subsection \ref{sub:cluster core evolution}.
In Subsection \ref{sub:TCE and TDE}, we show the BHs grow primarily through direct TDEs\footnote{Stars are destroyed at first pericentre passage.}. However, also
TCEs contribute to the BHs mass growth, as they trigger about $10$ per cent of the total TDEs.
We study how the TDE rate changes over time in Subsection \ref{sub:TDE rate}. There we provide fitting formulas and an analytical approximation to estimate
$\NdotTDE$, which also helps to understand quantitatively the dynamical processes that drive stars to disruption.
Finally, in Subsection \ref{sub:mass_growth}, we briefly discuss the phase of BH gas accretion after a star is tidally destroyed. Moreover, we investigate how the BH mass growth
changes varying the fraction of stellar mass that ends up into the BH after a TDE.

\begin{figure}
    \includegraphics[width=0.95\columnwidth]{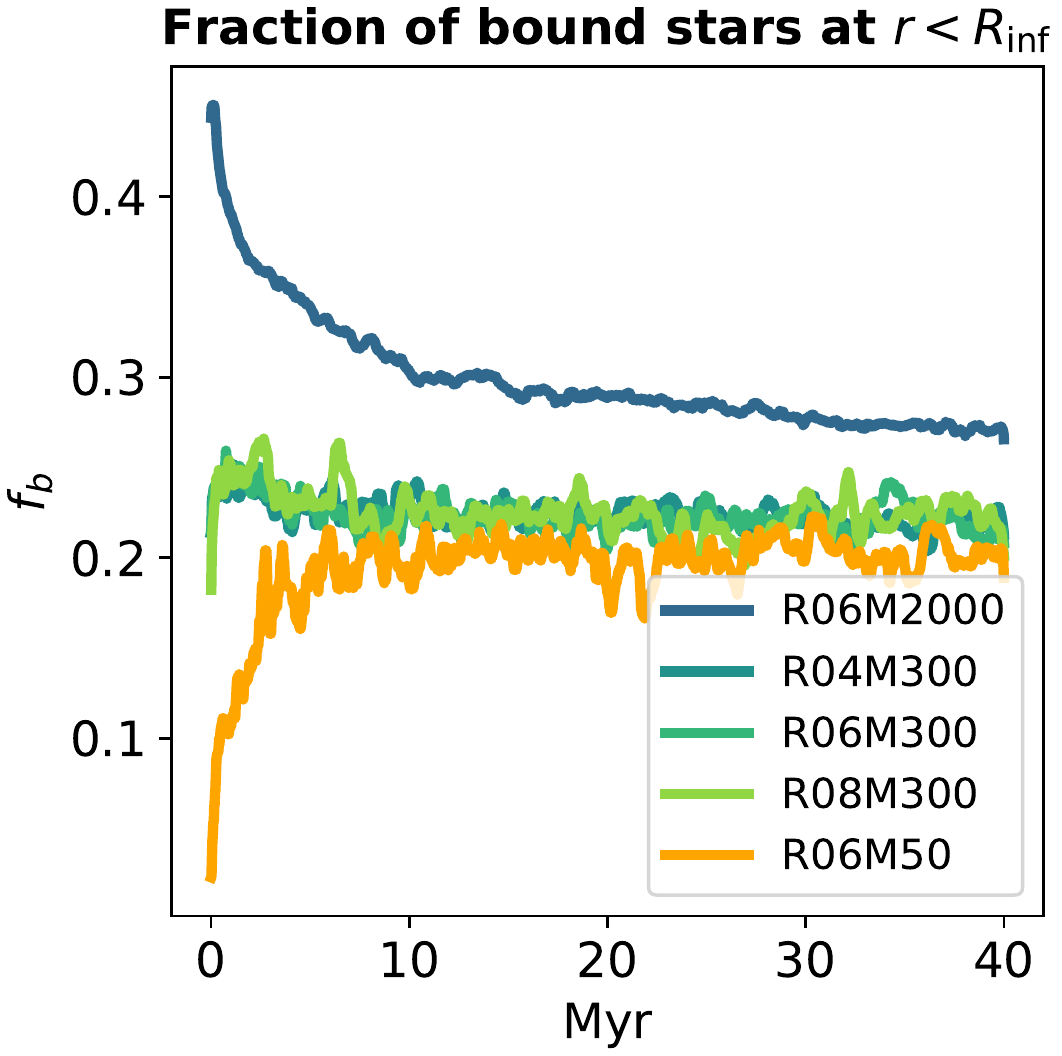}
    \caption{Time evolution of the fraction of stars $(f_{\mathrm{b}})$ within the BH influence radius ($\Rinf$) which are bound to the central BH.
    Models R04M300, R06M300, and R08M300 start with about $\sim25$ per cent of bound stars, and they maintain this fraction until the end of the simulations. R06M2000 starts with a fraction of about $\sim50$ per cent that drops to $\sim30$ per cent.
    A bound cloud of stars in R06M50 forms within the first 5 Myr reaching $\sim20$ per cent after $\sim 10$ Myr.}
    \label{fig:7_binary_fraction}
\end{figure}

\subsection{Formation and evolution of the bound cloud}\label{sub:bound_cloud}

Our simulations reveal that, in the vicinity of the BHs, inside a region enclosed within the influence radius $\Rinf$\footnote{The influence radius $\Rinf$ defines a sphere centred on the BH that encompasses a number of stars with total mass equal to $2\Mbh$.}, a significant fraction of the stars is bound to the BH i.e. they have negative orbital energy $E =  \frac{m_*v^2}{2} -\frac{Gm_*\Mbh}{r}<0$, where $m_*$ is the mass of the star. In Fig. \ref{fig:7_binary_fraction}, we plot, as a function of time, $f_{\mathrm{b}}$ the fraction of stars within the influence radius bound to the BH\footnote{By definition $f_{\mathrm{b}}=\frac{\Nb}{N_{\mathrm{inf}}}$, where $\Nb$ and $N_{\mathrm{inf}}$ are respectively the number of bound stars and the total number of stars within the influence radius.}.
All the models that started with a $\Mbh = 300 \mathrm{M_{\odot}} $ BH display a stable bound fraction approximately equal to $\approx0.2$. Fig. \ref{fig:7_binary_fraction} shows that this value stays constant for at least $40$ Myr. Since we evolved R06M300 for $\Tend$, we know that the bound fraction is very likely to remain constant for a significantly longer time (see Fig. \ref{fig:Appendix_R06M300} in Appendix \ref{apx:bound_cloud}).
On the other hand, model R06M2000 initially has a fraction of bound stars close to $50$ per cent, which drops rapidly in the first $\sim 10$ Myr and stabilises then at $30$ per cent, as indicated by the dark blue line in Fig. \ref{fig:7_binary_fraction}.  

In contrast to the other models, R06M50 starts its evolution without bound orbits around the BH. However, the bound cloud forms rapidly in the first $\sim 4$ Myr (orange line in Fig. \ref{fig:7_binary_fraction}). At the end of this build-up phase, the fraction of bound stars levels out at about $20$ per cent, a similar value to the one observed in R04M2000, R06M2000 and R08M2000. 

\begin{figure*}
    \includegraphics[width=0.98\textwidth]{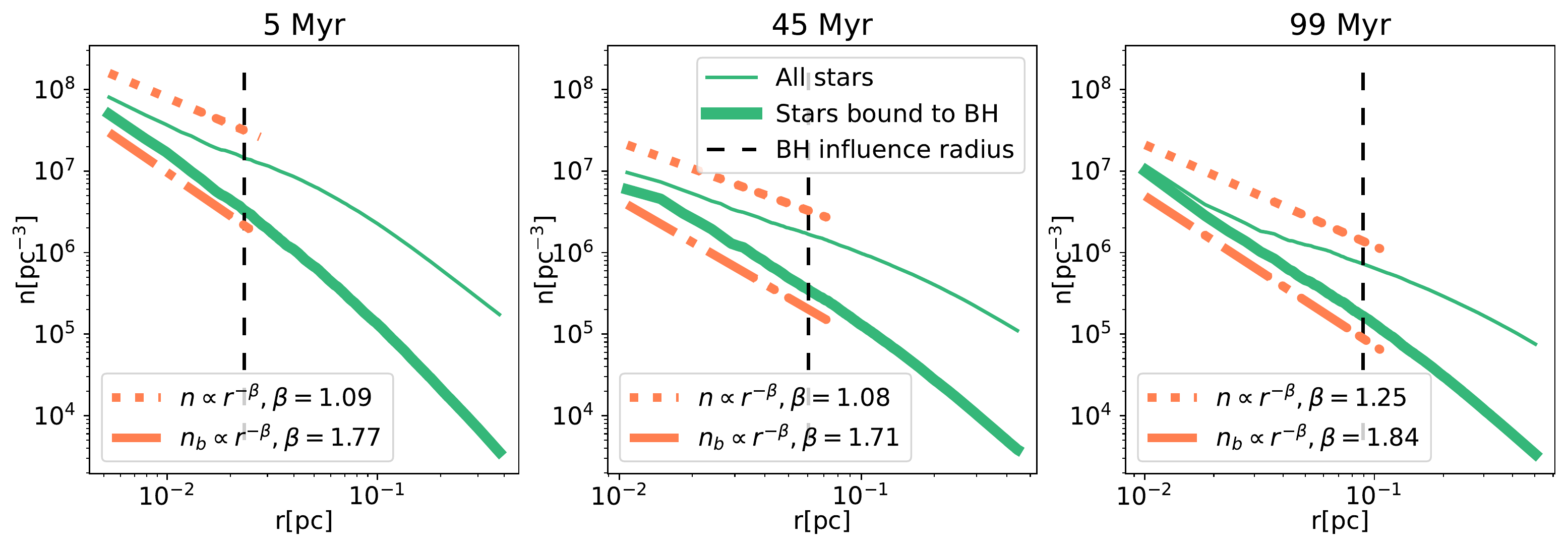}
    \caption{Total stellar number density profiles (thin green line) for the model R06M300 with a 300 $\mathrm{M_{\odot}} $ central BH at 5 Myr  (left), 45 Myr (middle) and 100 Myr (right). The power-law fits indicate shallow profiles with slopes of $\beta \sim 1.08 - 1.2$. The number density profiles of stars bound to the central black hole (bold green lines) are significantly steeper with power-law exponents of $\beta \sim 1.71 - 1.84$ (orange dot-dashed lines). These slopes are consistent with the \citet{Bahcall1976} $\beta = 7/4$ density profile (see Fig. \ref{fig:5_BW_comparison} for the time evolution of the power-law indices).
    }
    \label{fig:4_density_profile_fit}
\end{figure*}

\begin{figure*}
    \includegraphics[width=0.98\textwidth]{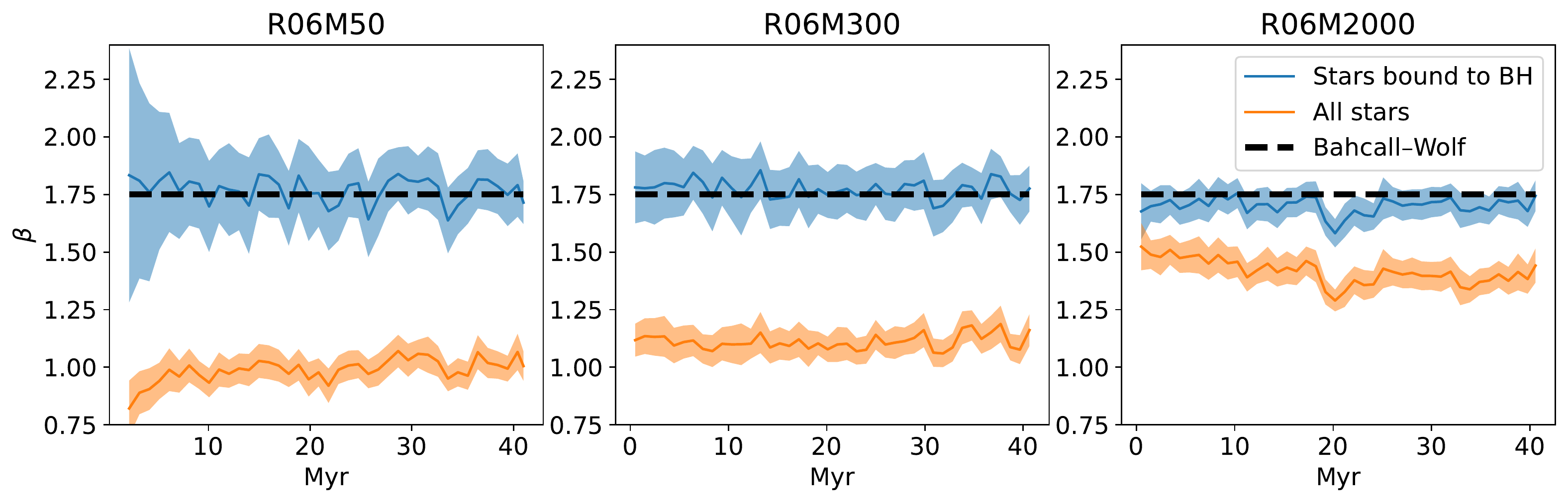}
    \caption{Time evolution of the power-law index $\beta$ for the stellar number density (see Fig. \ref{fig:4_density_profile_fit}) of all stars (orange) and stars bound to the central BH (blue). The central BH mass is increasing from $50 \mathrm{M_{\odot}} $ (R06M50, left) to 300 $\mathrm{M_{\odot}} $ (R06M300, middle) and 2000 $\mathrm{M_{\odot}} $ (R06M2000, right). The total density profile slope increases with increasing BH initial mass. It varies from $\beta \sim 1.0$ for R06M50 (left panel) to $\beta \sim 1.4$ for R06M2000 (right panel). The slope of the bound stars is independent of the black hole mass and mostly consistent with a Bahcall-Wolf cusp ($\beta \sim 1.75$). This result also holds for all other models and also on longer timescales (see Appendix \ref{apx:bound_cloud}). The uncertainty on the fitted slope, indicated in the plots with shadow regions, is estimated using one hundred consecutive simulation snapshots.}
    \label{fig:5_BW_comparison}
\end{figure*}

\subsection{Stellar distribution around the central black hole}
\label{sub:Bahcall Wolf}

When a single-mass stellar system with a massive central BH evolves for more than a relaxation time, gravitational interactions drive the stellar distribution in the vicinity of the BH to form a cusp with a particle density profile: $n \propto r^{-1.75}$, where $r$ is the distance from the BH and $n$ is the stellar particle density. 
This cusp, known as a Bahcall-Wolf (BW) cusp, was predicted for a stellar system with equal mass stars by \citet{Bahcall1976} 
and thereafter it was generalised for systems with unequal mass stars \citep{Bahcall1977, Alexander2009}.
In particular, \citet{Alexander2009} show that heavy stars, when they are relatively common, tend to follow the original BW profile with slope $\beta=1.75$, while lighter stars form a shallower cusp with $1.5\leq \beta \leq 1.75$.
Direct $N\mathrm{-body} $ simulations of star clusters with a central SMBH confirm that stars very close to the massive object follow a BW cusp and indicate that the cusp extends for a tenth of the influence radius $\sim 0.1 \Rinf$ \citep{Preto2004}.
Our runs show somewhat different results. The entire stellar population neighbouring the BH follows a significantly shallower distribution than the BW cusp. However, the distribution of the bound stars surrounding the BH is very much consistent with a BW profile that extend for about an influence radius $r<\Rinf$.   

To visualise the radial disposition of the stars around the BH in our simulations, we plot the radial stellar density for R06M300 at three different times, displayed with the green thin line in Fig. \ref{fig:4_density_profile_fit}.  
Within the influence radius, this  line is well fitted by a power-law $n \propto r^{-\beta}$ with slope that oscillates around $\beta\sim1.1$ (see middle panel of Fig. \ref{fig:5_BW_comparison}). On the other hand, the distribution of the bound component (green thick line in Fig.  \ref{fig:4_density_profile_fit}) is remarkably similar to the Bahcall-Wolf density profile. The relaxation timescale at $\Rinf$ for R06M300 is very short: gravitational scattering processes are expected to form a Bahcall-Wolf in less than $\lesssim 5$ Myr\footnote{Here the relaxation timescale is estimated accounting for both bound and unbound stars.}. However, only the bound orbits are affected by relaxation processes as they stay in the vicinity of the BH for a prolonged time. \\
This result holds for R06M50 and R06M300, as shown in the left and the central panels of Fig. \ref{fig:5_BW_comparison}, which illustrate that the power-law index $\beta$ for the bound cloud (in blue) oscillates around the Bahcall-Wolf predicted value throughout the simulation. 
Although the bound clouds in R06M50 and R06M300  consist of unequal mass stars, they are mainly composed of the heaviest stars in the cluster (due to primordial mass segregation), which, as predicted by \citet{Alexander2009}, follow a BW density profile.
This is a robust result that does not depend on the initial half-mass radius of the cluster (see Fig. \ref{fig:Appendix_BW}). \\
On the other hand, the power-law index $\beta$  of R06M2000 bound cloud is systematically lower than  $1.75$ throughout the simulation, as shown in the right panel of Fig.  \ref{fig:5_BW_comparison}. R06M2000, having a larger influence radius than R06M50 and R06M300, also has a more extended bound cloud containing many more low-mass stars. The latter tend to form a density profile shallower than the BW cusp as indicated by \citet{Alexander2009}.

\subsection{Evolution of the cluster core}
\label{sub:cluster core evolution}

\begin{figure}
    \includegraphics[width=0.95\columnwidth]{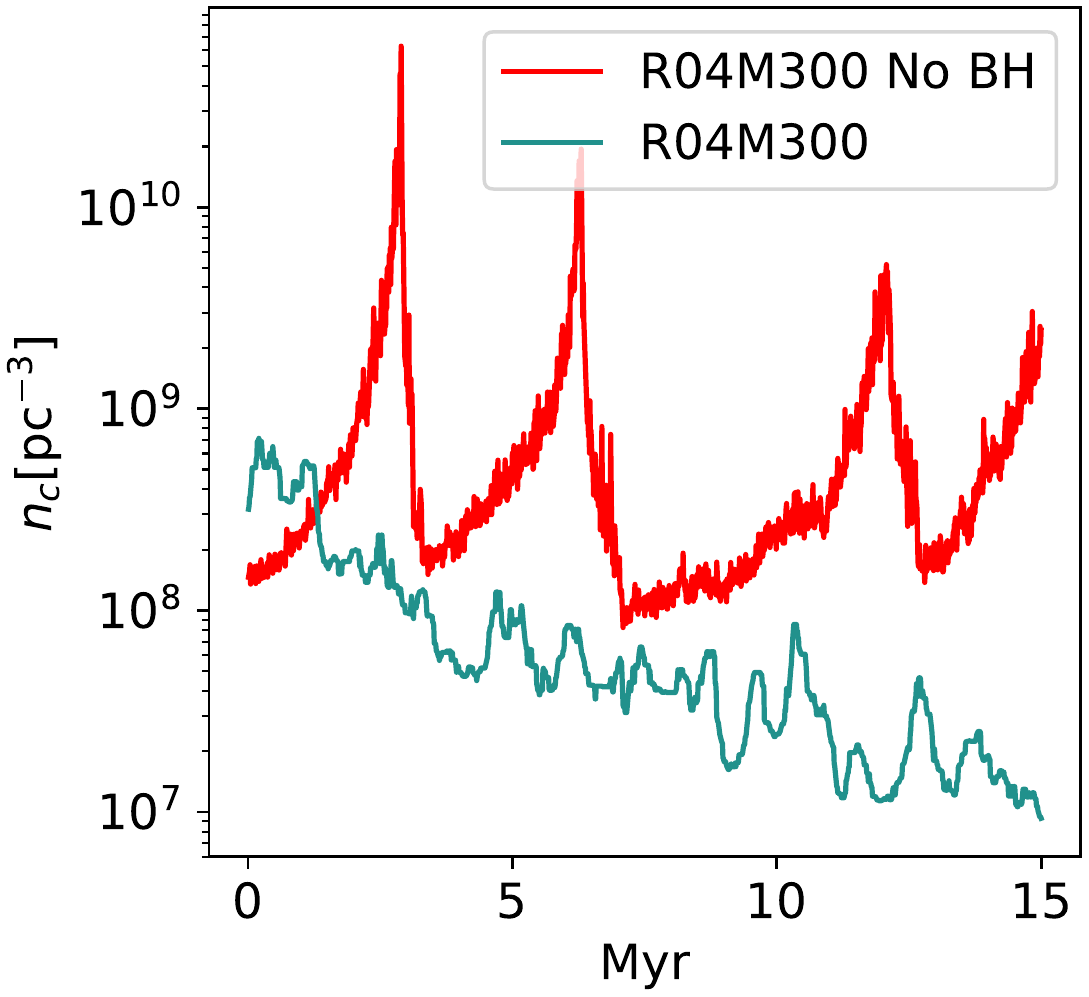}
    \caption{Time evolution of the central density (red) for the star cluster model R04 without a central black hole. The system rapidly undergoes core collapse and develops gravothermal oscillations driven by dynamically forming binary stars. The same cluster model with a central black hole of 300 $\mathrm{M_{\odot}} $ (R04M300, blue) does not undergo core collapse but expands gradually. The prevention of core collapse and the steady expansion is driven by the already existing population of "binaries" which the central black hole forms with many stars.}
    \label{fig:6_gravothermal_oscillations}
\end{figure}

The BHs in our runs constantly occupy the very centre of the cluster. Therefore,
their mass growth is heavily influenced by the cluster core properties (such as the density $\rho_\mathrm{c}$ and the velocity dispersion $\sigma_{\mathrm{c}}$) and their evolution over time.
At the same time, the BHs, exerting violent dynamical interactions in their surroundings, are expected to affect substantially the evolution of the cluster inner region.

To understand how the central compact object impacts the evolution of the cluster core, we rerun model R04M300 without a BH for 15 Myr.
Fig. \ref{fig:6_gravothermal_oscillations} compares the evolution of the R04M300 core density with and without a BH.
R04M300 without a BH (red line) undergoes gravothermal oscillations\footnote{ Gravothermal oscillations are well understood and have been studied extensively theoretically.
The first work that revealed such oscillations goes back to \citet{Sugimoto1983}. For a comprehensive overview of this topic see \citet[][]{Heggie1993}.}: its core collapses and expands several times.
Consequently, the central density oscillates from  $10^8$ up to $10^{11}$ particles per pc$^{3}$. The collapse phase is driven by two-body gravitational interactions, which randomly scatter particles into the centre, causing the core to contract \citep{Spitzer1987}. When the core is dense enough three-body, close interactions can efficiently form hard binaries, which release energy into the core, forcing it to expand. The expansion phase ends when the binaries escape the centre due to violent dynamical interactions.
Two-body interactions force the core to re-collapse as soon as all the binaries are ejected away from the centre, and the cycle restarts. Conversely, the core in R04M300 with a BH (blue line in Fig. \ref{fig:6_gravothermal_oscillations}) experiences a monotonic expansion. In this simulation, stars near the BH quickly bind to the compact object
BH quickly bind to the compact object and form BH-star binaries (many stars form a binary with the central BH), which
cause the core to expand. Due to the high BH mass, BH-stars binaries are unlikely to be ejected from the core. Instead, they can provide a steady energy flow to the centre until the end of the cluster evolution.
In other words, the central BH and the bound close stars act as a dynamical heat source and force the core to a steadily expand \citep[as already shown in previous works][]{Shapiro1977, Heggie2007}.
\begin{figure}
    \includegraphics[width=0.95\columnwidth]{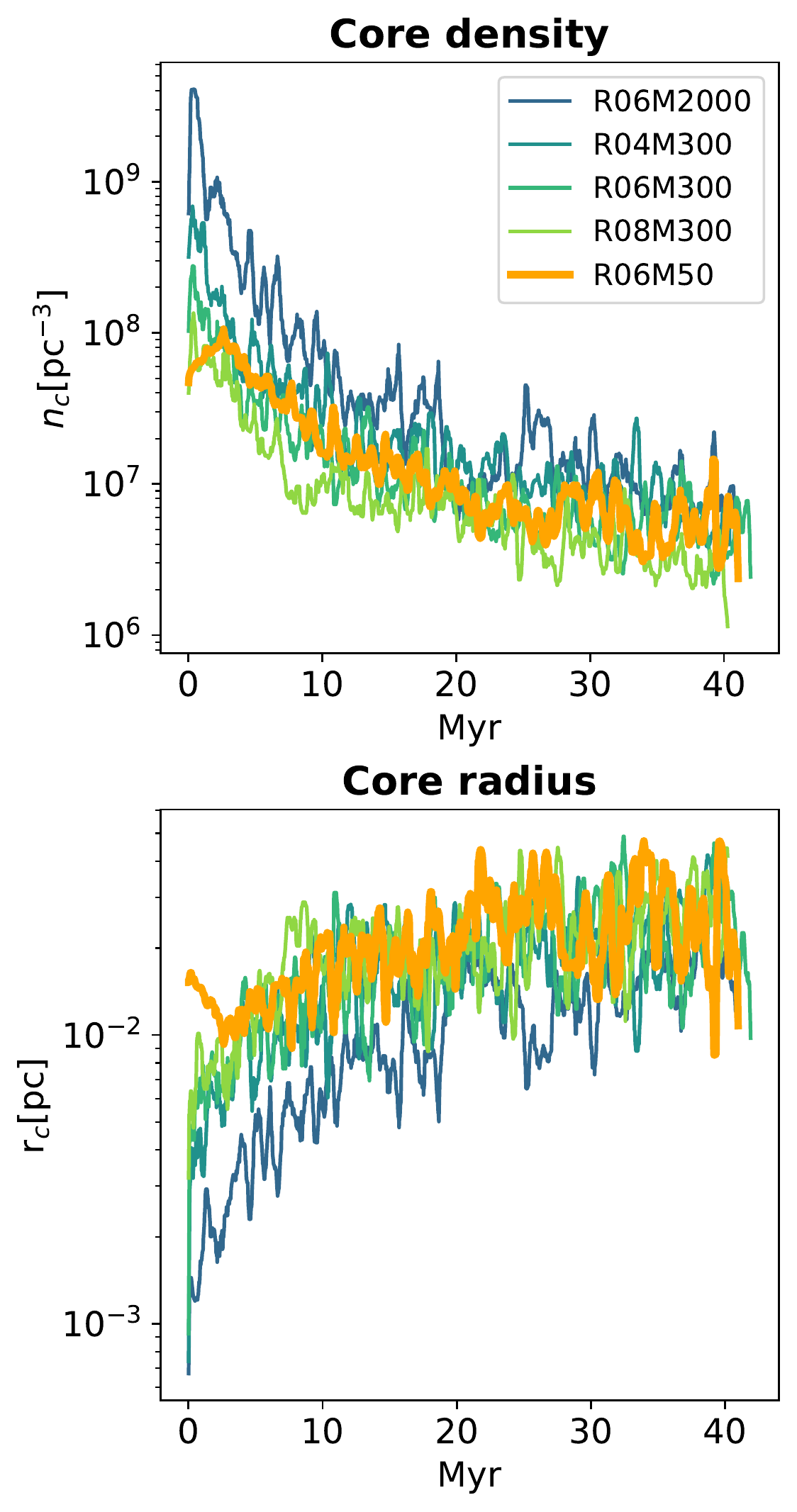}
        \caption{Time evolution of the central number density (top panel) and core radius (bottom panel) for the models as indicated in the legend. The initially very dense systems expand rapidly and the central density drops by almost two orders of magnitude. Stars bound to the central black holes drive the expansion. This effect prevents the expected core collapse in the absence of a central black hole (see Fig. \ref{fig:6_gravothermal_oscillations}). For the low mass (50 $\mathrm{M_{\odot}}$) black hole simulation (orange) the onset of core collapse is seen but is terminated once a bound nuclear subsystem has formed after a few million years (see Fig. \ref{fig:7_binary_fraction}).}
    \label{fig:8_core_expansion}
\end{figure}
Fig. \ref{fig:8_core_expansion} illustrates that R06M2000, R06M300 and R08M300 register early core expansion similar to R04M300. The expansion in these models starts right at the beginning of the simulation. On the other hand, in the first few million years, the core in R06M50 contracts (see orange line in Fig. \ref{fig:8_core_expansion}). The contraction halts after $\sim 4$ Myr, and the core expands until the simulation ends. The $50 \mathrm{M_{\odot}} $ BH in R06M50 cannot immediately reverse the core contraction because, at $t=0$ Myr, it does not have a bound cloud. The bound subsystem builds up in about $4$ Myr (orange line in Fig. \ref{fig:7_binary_fraction}) right before the core starts to expand (orange line in Fig. \ref{fig:8_core_expansion}).
This fact further indicates that the bound stars around the BH are responsible for the dynamical heating of the cluster centre.

\begin{figure}
    \includegraphics[width=0.95\columnwidth]{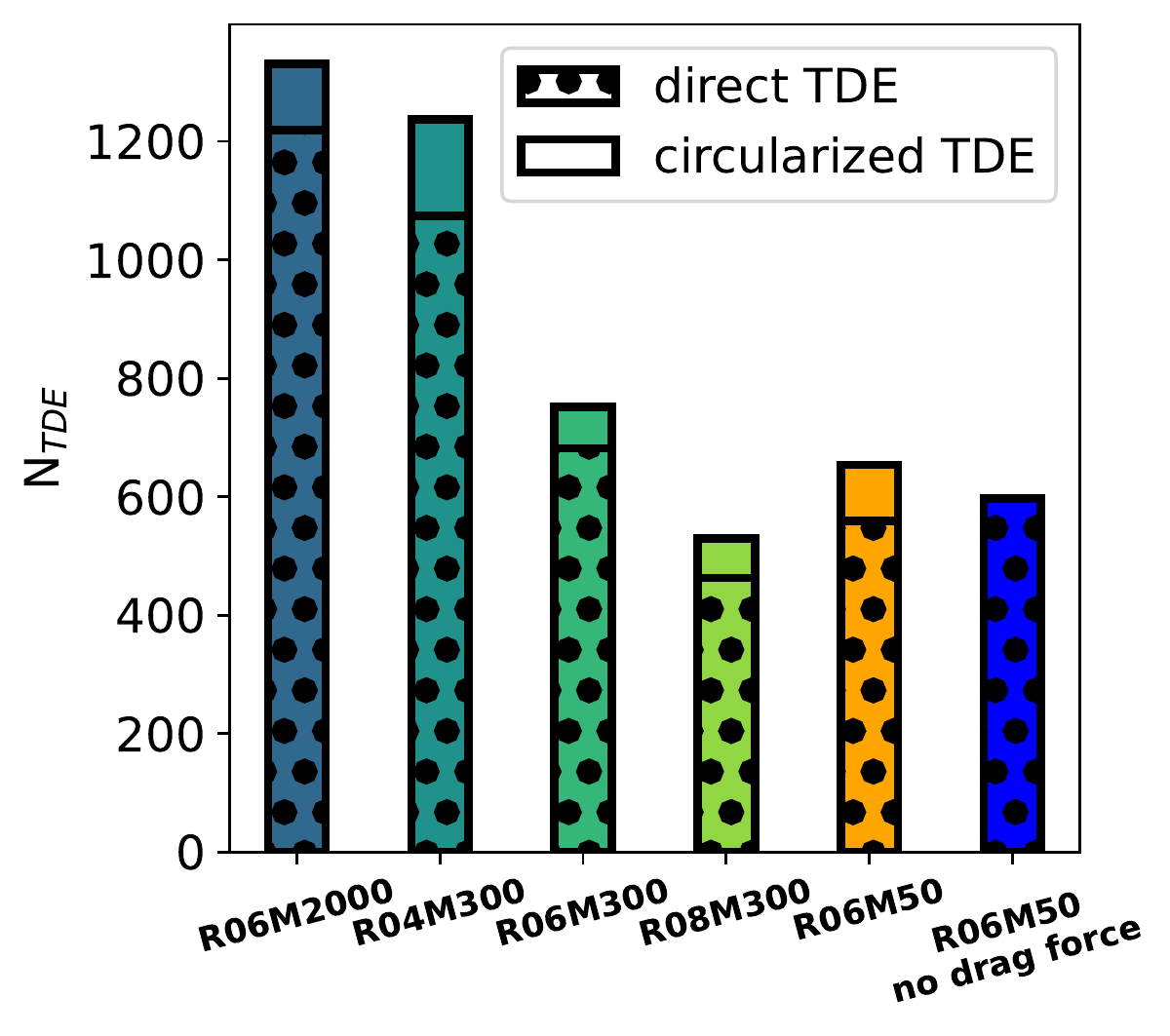}
    \caption{The number of tidal disruption events for the first 40 Myr of the simulations. The direct disruption events are dominating in all simulations (shaded regions) and the cicularised TDEs are sub-dominant. As explained in the text circularised TDEs correspond to TCEs. They contribute between $9$ per cent and up to $15$ per cent depending on the model. For comparison we show the R06M50 simulation without a tidal drag force (blue) for which the number of TDEs is reduced by about $10$ per cent. More massive central black holes (R06M2000) as well as more compact clusters (R04M300) increase the number of disruption events.}
    \label{fig:9_circ_and_direct_TDE}
\end{figure}

\begin{figure*}
    \includegraphics[width=0.98\textwidth]{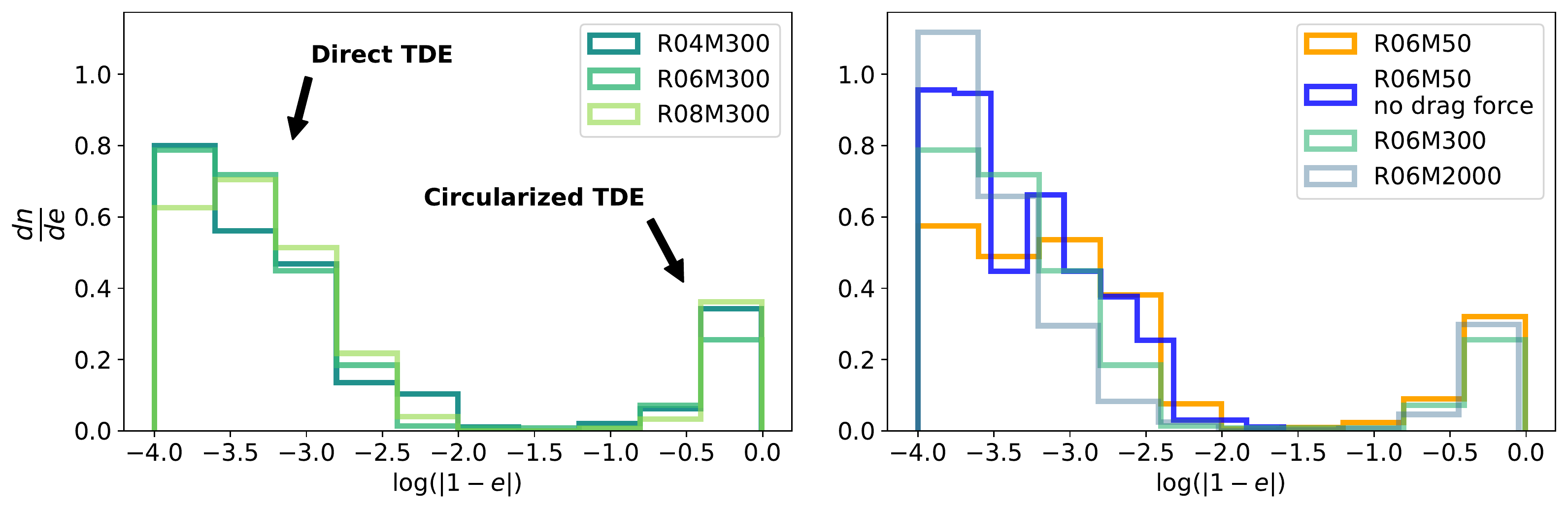}
    \caption{The eccentricity distribution of the stellar orbits right before tidal disruption for the M300 models with varying initial half mass radii (left panel) and the R06 models with varying central black hole masses (right panel). For all models most disruption orbits are very radial with high eccentricities $e \gtrsim 0.99$ (these events are indicated in the plot as direct TDEs). In a few cases, the distruption orbit is even hyperbolic (unbound)  with  $e \gtrsim 1.0$ (not shown here). About $\sim 10$ per cent of the stars have been tidally captured and have undergone partial or total circularisation by drag forces before disruption (these events are indicated as circularized TDEs and correspond, as explained in the text, to TCEs). The size of the star cluster and the mass of the central black holes only has a weak impact on the eccentricity distribution (see Section \ref{subsec:drag force}.) }
    \label{fig:10_ecc_distribution}
\end{figure*}

\subsection{Direct tidal disruption events and tidal capture events} \label{sub:direct TDE}
\label{sub:TCE and TDE}

Overall, the early expansion of the core registered in all our models does not prevent the BH from experiencing a large number of TDEs. Fig. \ref{fig:9_circ_and_direct_TDE} shows that only after 40 Myr two of our models (R04M300 and R06M2000) register more than $\gtrsim1200$ TDEs, and even our least dense system, R08M300, recorded more than $\gtrsim 500$ TDEs in the same period of time. 

We are interested in understanding how many of these TDEs were induced by TCEs. 
Unfortunately, as we have modeled tidal interactions with a drag force, directly integrated into the equations of motion, it is not possible to record TCEs directly. 
Nevertheless, we can deduce indirectly how many of the TDEs were triggered by TCEs by looking at the eccentricity of a star right before its disruption. 
With the prescription adopted in this study, every captured orbit undergoes a phase of circularisation followed by a phase of pericentre shrinking (see Subsection \ref{sub:orbit under drag force}).
Thus, a tidally captured star always experiences partial or total circularisation before being tidally disrupted. On the contrary, direct TDEs tend to have eccentricities very close to unity. 
The two panels in Fig. \ref{fig:10_ecc_distribution} display the eccentricity distribution of all TDEs.  The distributions are bimodal with a broad peak at $\log(1-e) \lesssim -2.0$ and a narrow peak that extends between  $\log(1-e) \sim -1.0$ and $\log(1-e) \sim 0.0$. Such plots provide a clear cut separation between direct TDEs ($e \sim 1.0$) and circularised TDEs ($e \lesssim 0.9$).

In principle, dynamical interactions could decrease the eccentricity of some direct TDEs. However, in practice, this never occurs. Only the tidal interaction drag force can significantly impact the eccentricity evolution of these events.  
For instance,  when we deactivate the tidal interaction drag force, the model R06M50 registered only direct TDEs (no events with $e \lesssim 0.9$ see blue line in Fig. \ref{fig:10_ecc_distribution}). With the drag force, R06M50 records 87 circularised TDEs. 
In summary, we can estimate the number of TCEs by counting the number of circularised TDEs ($e \lesssim 0.9$). 

Fig. \ref{fig:9_circ_and_direct_TDE} indicates that the BHs feed mainly on direct TDEs. However,
TCEs have also a relevant impact on the total number of TDEs: they trigger between $10$ per cent and $15$ per cent of the TDEs. Their contribution to the BH mass growth is even more substantial. Contrary to a direct TDE, in a TCE, the stellar material is tightly bound to the BH. Hence, the BH could absorb a larger fraction of the stellar material. We discuss this in more detail in Section \ref{sub:mass_growth}.


\begin{figure*}
    \includegraphics[width=0.98\textwidth]{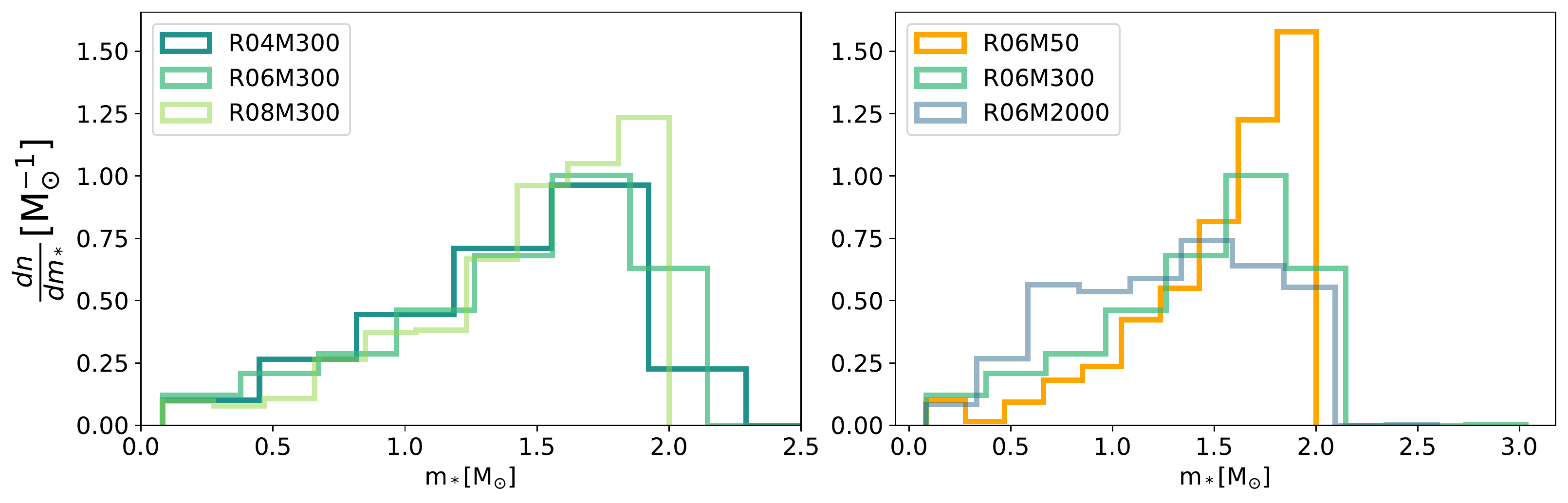}
    \caption{{\it Left panel}:  distribution of stellar masses $m_*$ tidally disrupted  (TDE) by the central BH for models R04M300, R06M300 and R08M300 (same initial BH mass and increasing half mass radii). Due to mass segregation the most massive stars in the simulations are more likely to be disrupted. Stars above $\sim 2$ $\mathrm{M_{\odot}}$ have formed in stellar mergers. {\it Right panel}: distribution of stellar masses $m_*$ tidally disrupted  by the central BH for models R06M50, R06M300 and R06M2000 (same initial half mass radius and increasing black hole masses). The distribution of R06M2000 looks markedly flatter  and does not peak at $\sim 1.7$ $\mathrm{M_{\odot}}$. This model has about three times more TDEs than the other models and the most massive stars are consumed in the first few million years. Thereafter increasingly lower mass stars are disrupted by the central black hole.} 
    \label{fig:11_star_mass}
\end{figure*}

\subsection{Tidal disruption rate as a function of time}
\label{sub:TDE rate}

\begin{figure*}
    \includegraphics[width=0.95\textwidth]{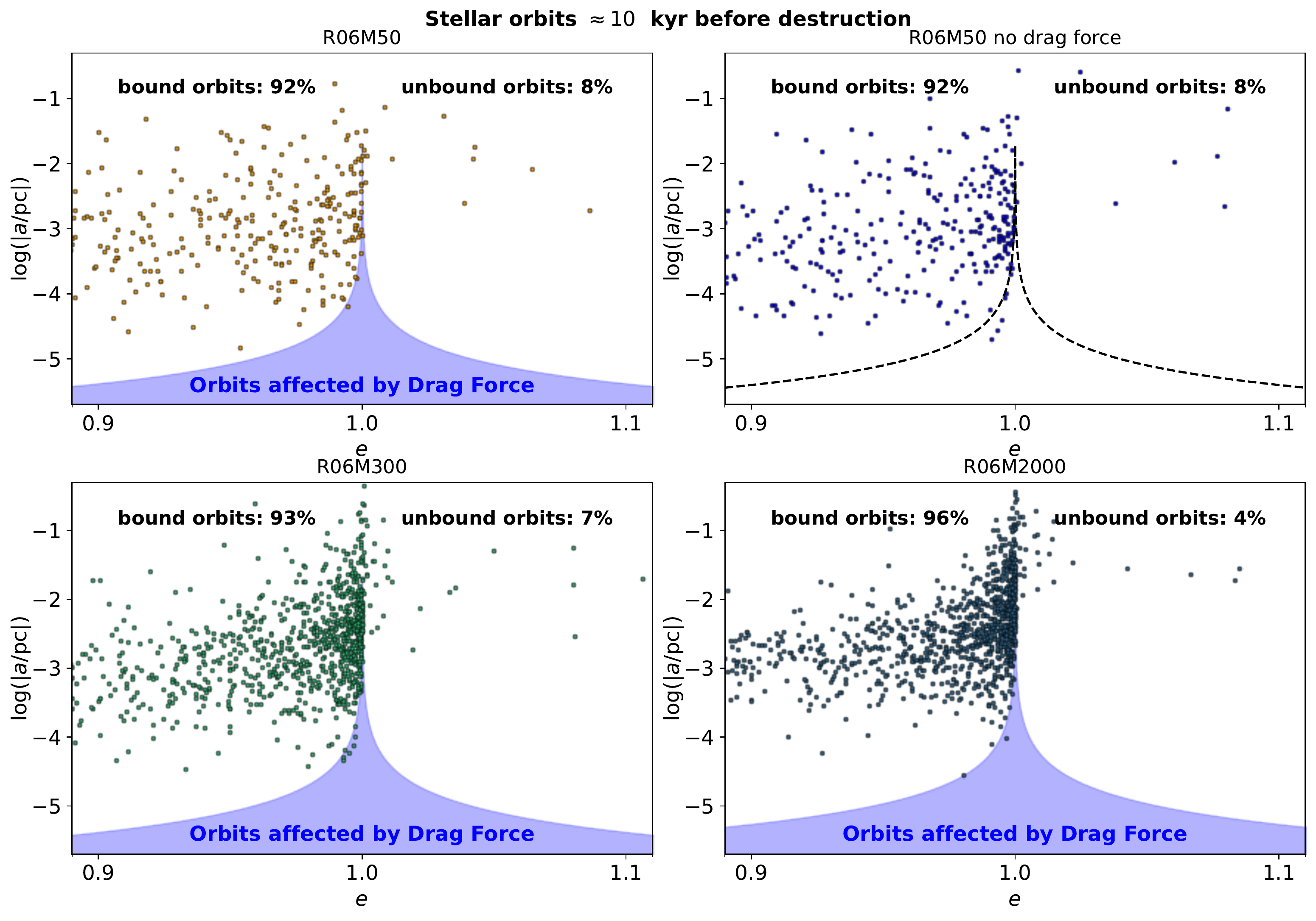}
    \caption{Semi-major axis vs. eccentricity about $10$ kyr before a star is disrupted by the central black hole for models R06M50, R06M300, and R06M2000 (from top left clockwise). The R06M50 model without a drag force is also shown for comparison at top right. The colored circles reported in each panel represent all the stars that merged with the BH. The orbital elements are computed before they could have been affected by tidal energy loss (here indicated with the blue area). More than $90$ per cent of the colliding stars were moving on a bound orbit.}
    \label{fig:13_TC_orbit_elemetnts}
\end{figure*}

Unsurprisingly, in our simulations, the average mass of the tidally disrupted stars is significantly higher than the mean stellar mass;
it varies from $m_*\sim 1.3$ $\mathrm{M_{\odot}}$ up to $m_*\sim 1.6$ $\mathrm{M_{\odot}}$  depending on the initial conditions.
In fact, due to mass segregation, the BH preferentially devours the most massive stars in the system.
The mass distribution of these stars is shown in Fig. \ref{fig:11_star_mass}.
For four of our models, the distribution peaks at $2 \mathrm{M_{\odot}}$, which are the most massive stars of the clusters. 
R06M2000 is the only exception; it has a rather flat mass distribution. The $2000 \mathrm{M_{\odot}}$ BH destroys with the same probability stars between  $\sim 0.5 \mathrm{M_{\odot}}$  and  $2.0 \mathrm{M_{\odot}}$.     
This exception can be explained by looking at Fig. \ref{fig:13_TC_orbit_elemetnts}, which shows that most TDEs originate from bound orbits.
Most of the stars that contribute to the growth of the BH come from the bound cloud. 
In other words, the BH feeds mainly on the bound subsystem. Due to its large BH mass, R06M2000 is the model with the most extended bound cloud since bound clouds extend for about an influence radius ($r<\Rinf$) (see Subsection \ref{sub:bound_cloud}). Therefore, compared to the other models, its bound subsystems encompass many more low-mass stars. To recap, the reservoir of stars that sustain the growth of the BH coincides mostly with the bound subsystem. Since R06M300 has the most extended bound cloud,   it has a higher probability of consuming lighter stars.

\begin{figure*}
    \includegraphics[width=0.98\textwidth]{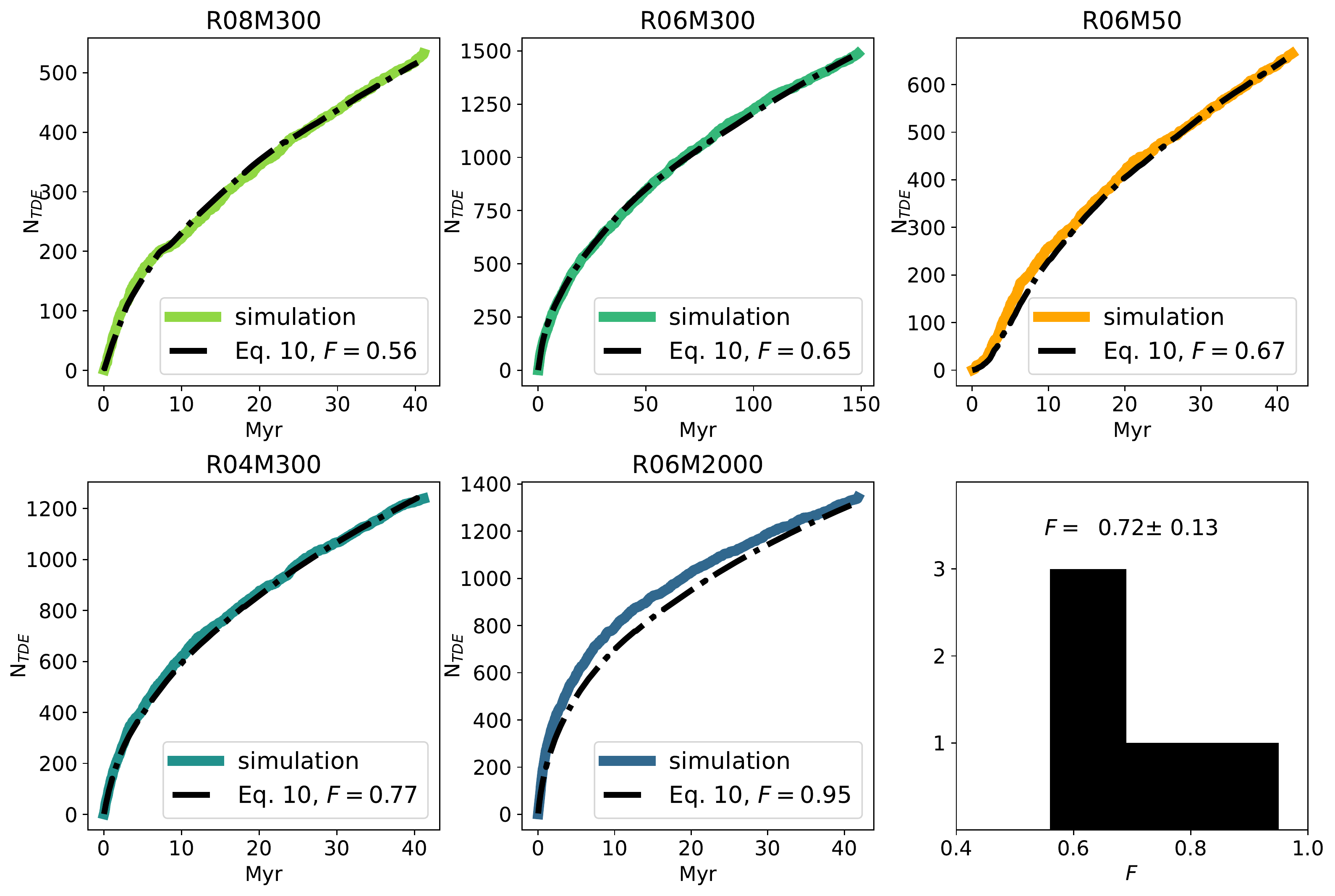}
    \caption{Cumulative number of TDEs fitted using equation \ref{eq:NTDE} for all the five models. After calibrating the constant factor $F$ for each model, Eq. \ref{eq:NTDE} predicts very accurately the number of TDEs as a function of time. The histogram in the right bottom panel reports the distribution of the $F$-values, as well as $F$ mean value $\pm$ the standard deviation. The evolution of the cumulative number of TDEs is displayed up to 40 Myr for all models, except for the R06M300 model which we evolved for $\Tend$. }
    \label{fig:15_r_power_law}
\end{figure*}

As we have already mentioned, the bound cloud plays a key role in driving most of the TDEs. Equally important is the role played by the unbound component of the cluster, which, first of all, acts as a reservoir for the bound subsystem, ensuring that the fraction of bound stars is constant over time (see Fig. \ref{fig:7_binary_fraction}). 
In addition, unbound stars constantly scatter with the bound orbits, and through complex dynamical interactions, they also trigger TDEs and TCEs. 
Note that orbits in the bound cloud are typically too wide to be affected by the tidal drag force: most of the orbits are beyond the drag force region of influence (blue area in Fig. \ref{fig:13_TC_orbit_elemetnts} ). Gravitational scattering events are necessary to lead stars close enough to be tidally captured or tidally destroyed. In other words, the two-body relaxation process is indispensable for driving the bound cloud stars to tidal disruption. 
The time scale at which orbits near the BH change their energy and angular momentum is consistent with the non-resonant relaxation time scale $\trel$. Since we include post-Newtonian corrections in the equation of motion of the simulation particles, resonant effects are suppressed in agreement with what previous studies have found \citep{Merritt2011, Brem2013}.
From these considerations it follows that we can estimate the TDE rate using:
\begin{equation}
    \label{eq:Ndot_1}
    \NdotTDE = F\frac{N_{b}}{\trel}
\end{equation}
where $N_{b}$ is the number of bound stars within the BH influence radius, $F$ is a numerical pre-factor that is calibrated for the different models and $\trel$ is the relaxation time scale of the cluster calculated at $\Rinf$ which we estimate following \citet{Spitzer1987}:
\begin{equation}
    \label{eq:t_rel}
    \trel = \frac{1.8\times 10^3}{\ln(\Lambda)} \left( \frac{10^7 \mathrm{M_{\odot}}/\mathrm{pc^3}}{\rho} \right) \left(\frac{\sigma}{100 \mathrm{km/s} } \right)^3  \left(\frac{1\mathrm{M_{\odot}}}{m_*} \right) \mathrm{Myr}
\end{equation}
where $\sigma$, $\rho$ and $m_*$ are respectively the velocity dispersion, stellar density and the average stellar mass computed within $r<\Rinf$.
The Coulomb logarithm, $\Lambda$, can be approximated with $\Lambda = 0.11 N$ \citep{Giersz1994}.  In our case $N$ is the number of particles within the influence radius, therefore $N=2 \Mbh / m_*$. Similarly we can now rewrite $N_{b}=2 f_{\mathrm{b}} \Mbh/m_*$, where $f_{\mathrm{b}}$ is the fraction of bound stars at $r<\Rinf$. It then follows that we can rewrite Eq. \ref{eq:Ndot_1} as: 
\begin{equation}
    \label{eq:Ndot_2}
    \begin{aligned}
    \NdotTDE = & 1.1 F f_{\mathrm{b}} \ln \left( 0.22 \frac{\Mbh}{m_*} \right) \left( \frac{\Mbh}{10^3 \mathrm{M_{\odot}}}\right)\left( \frac{\rho}{10^7 \mathrm{M_{\odot}}/\mathrm{pc^3}} \right) \\
    & \times \left(\frac{100 \mathrm{km/s} }{\sigma} \right)^3 \mathrm{Myr^{-1}} \ 
    \end{aligned}
\end{equation}
where the stellar velocity dispersion $\sigma$, the stellar density $\rho$ and the average stellar mass $m_*$ are computed at the influence radius $(r=\Rinf)$.
To estimate the cumulative number of TDEs as a function of time
we integrate Eq. \ref{eq:Ndot_2} obtaining:
\begin{equation}
    \label{eq:NTDE}
    \begin{aligned}
    \NTDE(t) =  1.1 F   \int_{0}^{t}& f_{\mathrm{b}}(t')\left( \frac{\Mbh(t')}{10^3 \mathrm{M_{\odot}}}\right) \ln \left( 0.22 \frac{\Mbh(t')}{m_*(t')} \right) \\
    & \times \left( \frac{\rho(t')}{10^7 \mathrm{M_{\odot}}/\mathrm{pc^3}} \right) \left(\frac{100 \mathrm{km/s} }{\sigma(t')} \right)^3 dt'. \
    \end{aligned}
\end{equation}
Here, to stay as general as possible, we assumed that the variables $f_{\mathrm{b}}$, $\Mbh$, $\rho$, $m_*$ and $\sigma$ functions of time.
Calibrating the variable $F$ on each model Eq. \ref{eq:NTDE} predicts fairly accurately the number of TDEs as a function of time as displayed in Fig. \ref{fig:15_r_power_law}. The bottom right panel of the same figure shows the best fitting value for $F$ in the different simulations and it indicates that $0.6\lesssim F \lesssim 1.0$.

\begin{figure}
    \includegraphics[width=0.95\columnwidth]{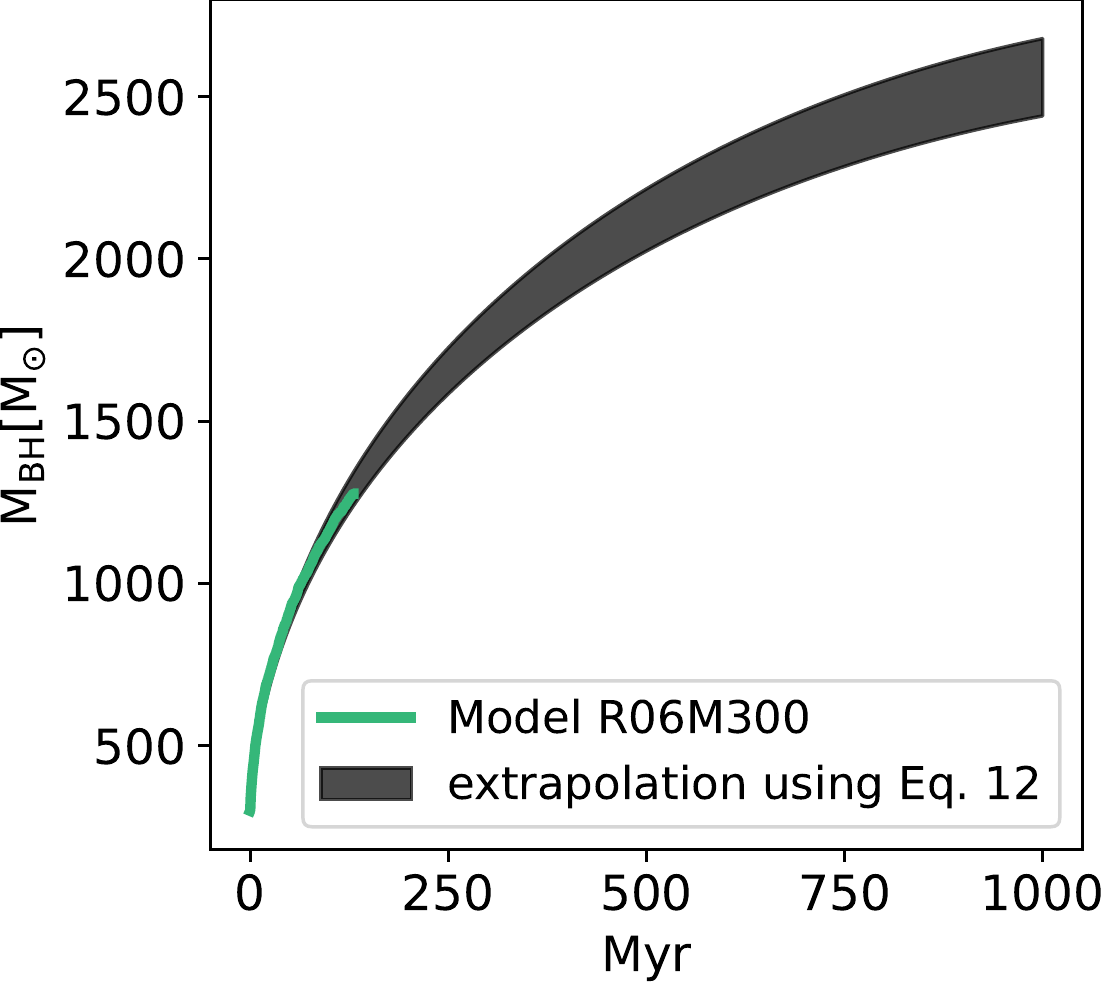}
    \caption{The mass of the BH as a function of time for the simulation R06M300 (green line) and using equation \ref{eq:Mdot} (black region). The black area perfectly reproduces the simulation result until $\Tend$ (the end of the simulation). It then predicts $\Mbh$ for later times.  
    The uncertainty in the prediction is calculated including the fitting error for $\rho$ and $m_*$ (see Subsection \ref{sub:mass_growth} for more details).}
    \label{fig:16_extrap_comparison}
\end{figure}

\begin{figure*}
    \includegraphics[width=0.95\textwidth]{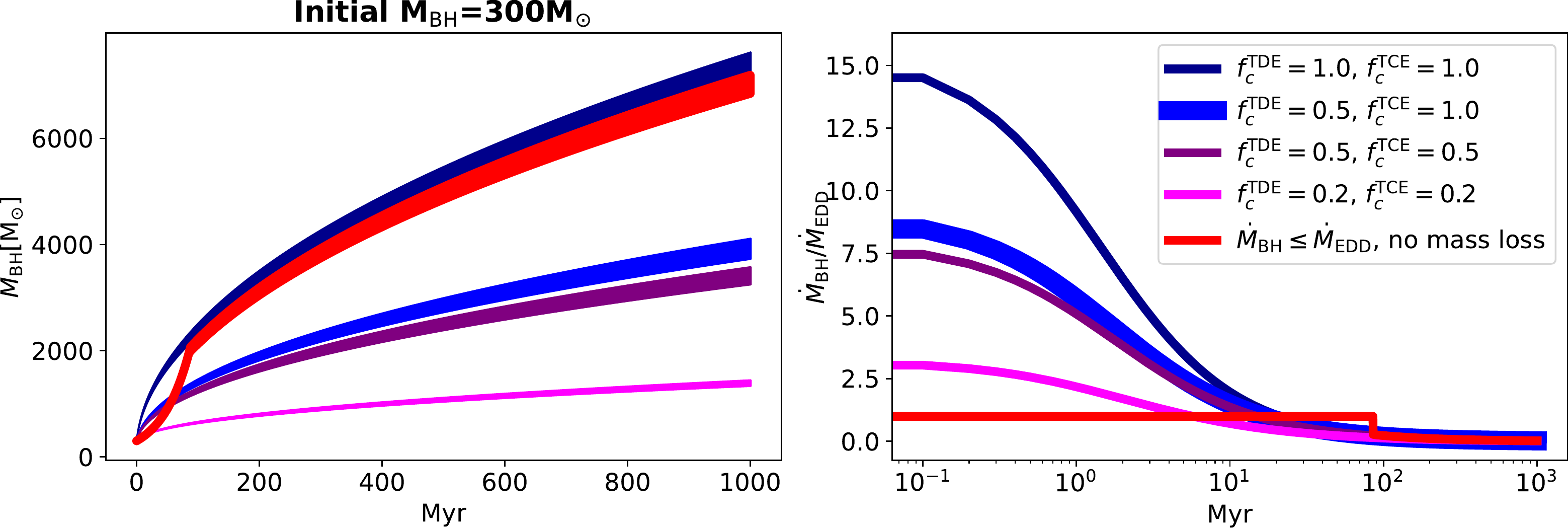}
    \caption{\textit{Left panel}:Mass of the BH as a function of time computed by integrating Eq. \ref{eq:Mdot}. When solving the equation we assume $m_*$ to be constant and equal to $m_*=1.5\mathrm{M_{\odot}}$; We vary $f_{\mathrm{c}}$ from $1.0$ (full accretion scenario, see dark blue region) to $0.2$ ($80$ per cent stellar material is lost, see magenta region). \textit{Right panel}: BH mass accretion as a function of time for the five scenarios presented in the left panel. In one scenario (red line) we limit the accretion to the Eddington rate and we assume all the stellar materia in TDEs accumulate around the BH without mass loss. }
    \label{fig:17_extrapolating_BH_growth}
\end{figure*}

\subsection{Extrapolating the mass growth of the central black hole}
\label{sub:mass_growth}
This subsection describes a method to predict the BH mass growth beyond the simulation time.  We cannot evolve our systems for more than $\Tend$ as our simulations are computationally costly, but we can use the results of our runs to extrapolate the BH mass for a longer time.
We start by connecting the tidal disruption rate $\NdotTDE$ estimated in equation \ref{eq:Ndot_2} with $\Mdotbh$ the BH accretion rate. 
For this purpose, we need to estimate how much stellar mass remains bound to the BH during a TDE and how much of the bound material ends up in the compact object.

The simple argument presented in \citet{Rees1988} indicates that if a BH destroys a star in a parabolic orbit, half of the stellar mass is lost during the disruption process.
The remaining portion is expected to form an accretion disk around the BH, eventually falling on the compact object on a time scale that depends on the accretion process \citep{Shiokawa2015, Bonnerot2016, Hayasaki2016}.  If the accretion disk injects gas into the BH at a rate larger than the Eddington rate, the disc likely becomes weakly bound
and a significant fraction of the gas is ejected through winds \citep{Strubbe2011}. 

In our simulations, we model neither the accretion disc formation phase nor the actual accretion phase. Instead we average out all these processes by introducing the parameter $f_{\mathrm{c}}$, which 
represent the fraction of stellar mass accreted by the BH during a TDE. It follows from the definitions of  $f_{\mathrm{c}}$ that:
\begin{equation}
\begin{aligned}
\Mdotbh = f_{\mathrm{c}} m_* \NdotTDE. 
\end{aligned}
\end{equation}
We can therefore estimate the BH mass growth rate $\Mdotbh$ by computing $\NdotTDE$ using Eq. \ref{eq:Ndot_2}.
Observing that the velocity dispersion to estimate $\NdotTDE$ is calculated at $r=\Rinf$ we can further simplify the previous equations by rewriting $\sigma$ as a function of the BH mass $\Mbh$ and the density $\rho$ using the virial theorem (see Eq. \ref{eq:sigmafit}). It thus follows:
\begin{equation}\label{eq:Mdot}
\Mdotbh = C f_{\mathrm{c}} f_{\mathrm{b}} \left( \frac{m_*}{\mathrm{M_{\odot}}} \right) \sqrt{ \frac{\rho}{10^7\mathrm{M_{\odot}}/\mathrm{pc}^3} } \ln \left( 0.22 \frac{\Mbh}{m_*} \right) \frac{\mathrm{M_{\odot}}}{\mathrm{Myr}}
\end{equation} 
where we have incorporated all the constants in $C=1.1 \times 10^6 F / C_0$ (see Appendix \ref{apx:extrapolations} for the definition of $C_0$).
Since our simulations are computationally very expensive, we evolved them for no more than $\Tend$. However, as long as we find a reliable extrapolation of  $f_{\mathrm{b}}$, $\rho$ and $m_*$, we can use Eq.  \ref{eq:Mdot} to predict the mass of the black hole beyond our simulation time.
It turns out that all three of these quantities, the density, the average stellar mass and the fraction of bound stars, manifest regular and predictable behaviour.
As transpires from the left panel of Fig. \ref{fig:Appendix_R06M300} it is reasonable to assume $f_{\mathrm{b}}$ does not change over time. Hence we consider $f_{\mathrm{b}} \approx 0.2$ to be constant.
The density $\rho$ in the vicinity of the BH (upper panel in Fig. \ref{fig:8_core_expansion}) displays a consistent trend: after an initial phase of core expansion,  all models, independently of their initial BH mass and initial concentration, converge to an almost identical evolution. Consequently, we can use a fitting function to robustly extrapolate the later evolution of  $\rho$, as we illustrate comprehensively in Appendix \ref{apx:extrapolations}. In the same appendix, we also show that the change in time of $m_*$ can be modelled using a linear fit (see the left panel in Fig. \ref{fig:Appendix_extrapolations}).

To summarise, we provided fitting formulas to predict the evolution of  $\rho$ and  $m_*$, which, together with Eq. \ref{eq:Mdot} allow us to extrapolate $\Mbh$ as a function of time. 
Substituting in Eq. \ref{eq:Mdot} $f_{\mathrm{c}}=0.5$,  which is the fiducial value we assumed in our runs, we obtain the results displayed in Fig.  \ref{fig:16_extrap_comparison}.
The time evolution of the R06M300 BH mass  is perfectly consistent with the simulation results (green line in Fig.  \ref{fig:16_extrap_comparison})
reaching $2500 \mathrm{M_{\odot}} $ in about $\sim 1$ Gyr.

Thanks to its simplicity, we can use Eq. \ref{eq:Mdot} to explore systems that are slightly different from the ones we simulated. For instance, a star cluster with a larger number of particles ($N>10^6$) has a larger number of massive stars at its disposal that the BH can consume. This implies that $m_*$ stays constant for much longer. Simply substituting the star's mass to a constant value of $m_*=1.5 \mathrm{M_{\odot}}$, leads to a more significant mass growth: the BH in this scenario is expected to reach $3500 \mathrm{M_{\odot}}$ in 1 Gyr (see the purple line in Fig. \ref{fig:17_extrapolating_BH_growth}).
By means of equation \ref{eq:Mdot} we can also explore how modifying $f_{\mathrm{c}}$ would affect our findings. We are not only limited to changing the value of $f_{\mathrm{c}}$  but we can also make a distinction between the accretion fraction for direct TDEs ($f_\mathrm{c}^\mathrm{TDE}$) and for TCEs  ($\fcTCE$).
Since our models registered $\sim 90$ per cent of direct TDEs and about $\sim10$ per cent of TDEs triggered by tidal capture. We can thus link $f_{\mathrm{c}}$ with $f_\mathrm{c}^\mathrm{TDE}$ and $\fcTCE$ as follow:
\begin{equation}
f_{\mathrm{c}} = (0.9f_\mathrm{c}^\mathrm{TDE} + 0.1\fcTCE) / 2.
\end{equation}
We use the equality above to explore the scenario
$\fcTCE =1.0, f_\mathrm{c}^\mathrm{TDE} =0.5$, which is motivated by the fact that, in TDEs induced by 
tidal captures, stellar gas is more tightly bound to the BH.
The left panel of Fig. \ref{fig:17_extrapolating_BH_growth} outlines the main results.
Not surprisingly $f_\mathrm{c}^\mathrm{TDE} = \fcTCE =0.2$ fades the growth of the BH, which reaches only $\sim 1500 \mathrm{M_{\odot}}$ in 1 Gyr.  On the other hand,  the scenario with $f_\mathrm{c}^\mathrm{TDE} = \fcTCE =1.0$ brings the BH to $\sim 7600 \mathrm{M_{\odot}}$.  This scenario might require an initial accretion rate of $\sim 15 \dot{M}_{\mathrm{EDD}}$ as the dark blue line in the right panel of Fig. \ref{fig:17_extrapolating_BH_growth} indicates. Observations indicate that super-Eddington accretion
can be possible during TDEs \citep[see][and references therein]{Komossa2015}. 
However, theoretical studies show that jets and violent outflows 
will probably unbind between $30$ per cent and up to $75$ per cent of the gas around the BH \citep{Ayal2000, Metzger2016, Toyouchi2021}. For this reason, the BH in our model might not be able to sustain an accretion rate with $f_{\mathrm{c}}=1$, at least during the first $20 - 30$ Myr.  
Nevertheless, the BH can still reach $\sim 7000 \mathrm{M_{\odot}} $ in 1 Gyr, even limiting the accretion rate to the Eddington accretion as shown (in red) in the left panel of Fig. \ref{fig:17_extrapolating_BH_growth}. If all the stellar gas generated in TDEs accumulates around the BH and stays bound to it, the BH will accrete it within 1 Gyr, even restricting its growth rate to $\Mdotbh \lesssim \dot{M}_{\mathrm{EDD}}$. The accretion rate for this scenario, indicated in the right panel of Fig. \ref{fig:17_extrapolating_BH_growth} in red, is constant $\Mdotbh=\dot{M}_{\mathrm{EDD}}$ for the first $\sim 100$ Myr; as soon as the BH consumes all the gas accumulated during the past TDEs, the accretion drops rapidly to significantly lower values.  
In conclusion, our analysis indicates that $300 \mathrm{M_{\odot}} $ BH can grow up to $\sim 10^4 \mathrm{M_{\odot}} $ as long as, after TDEs, all stellar material remains bound. This scenario might be unfeasible if the tidally destroyed stars move in open orbits, but as we showed in the previous subsection that the BH feeds primarily on the bound cloud. To put it in another way, the great majority of the destroyed stars come from bound orbits. Therefore, most stellar material formed during TDEs might have a high chance of remaining bound to the BH.

\section{Discussion and conclusions}
\label{sec:Conclusions}
In this work, we evolved and analyzed five direct $N\mathrm{-body} $ simulations of compact ($\Rh\leq0.8$ pc) star clusters ($N=256000$) for at most $\Tend$ to investigate the growth of massive BHs through repeated TCEs and TDEs. All our systems start with a single central BHs ($50 \mathrm{M_{\odot}} \leq \Mbh \leq  2000 \mathrm{M_{\odot}} $) surrounded by mass-segregated low-mass stars in the mass range $0.08 \mathrm{M_{\odot}} \lesssim m_*\lesssim2.0 \mathrm{M_{\odot}}$. 
The clusters initial central stellar density $n_{\mathrm{c}} > 10^7 $pc$^{-3}$
and initial central velocity dispersion ($30 \mathrm{km/s} \lesssim \sigma_{\mathrm{c}} \lesssim 90 \mathrm{km/s}$) fulfill the criteria provided by \citet{Stone2017} to trigger tidal capture runaway collisions.
The five models show several similarities in both the evolution of the cluster core and the growth of the central black hole.
The BHs in all realizations, at the very beginning of the runs, give rise to a subsystem of bound orbits in its vicinity. Around $20$ per cent to $30$ per cent of the stars enclosed within the influence radius ($r<\Rinf$) stays bound to the BH throughout the simulation. This cloud of bound stars forms a \citet{Bahcall1976}  cusp embedded in a reservoir of unbound particles with a shallower density distribution. 
Our analysis indicate that the bound cloud plays a key role in the evolution of the cluster core, as it constantly releases energy to the core through dynamical interactions resulting in a monotonic core expansion.  
Consequently, the clusters cannot mantain their original central density which drop by about an order of magnitude in the first few Myr.
Despite this initial expansion, the BHs experienced a sustained tidal disruption rate until the simulation ended. About $10-15$ per cent of the TDEs are caused by tidal captures; in all these cases, the orbit of the destroyed stars underwent partial or total circularisation. \\
We derived simple equations (see Eq. \ref{eq:Ndot_2} and \ref{eq:NTDE}) that predict accurately the number of TDEs experienced by the central BHs as shown in Fig. \ref{fig:15_r_power_law}. 
We use these equations to derive an expression that model the BH mass growth rate as a function of time.
With this expression we explore scenarios that slightly deviate from our initial conditions and predict the time evolution of BH masses up to 1 Gyr. 
We predict that a few hundred solar masses BH can easily reach $\sim 2000 - 3000 \mathrm{M_{\odot}}$ through repeated TDEs and TCEs within 1 Gyr. The BH
can even grow up to $\sim 10^4\mathrm{M_{\odot}}$ if all the stellar material remains bound to the BH and no mass loss occur during the destruction and accretion phase (see Fig. \ref{fig:17_extrapolating_BH_growth}). We argue that this scenario is plausible because, as revealed by our simulations (see Fig. \ref{fig:13_TC_orbit_elemetnts}), the BHs feed primarily on bound orbits. As long as the accretion proceeds at sub-Eddington rate, all the stellar gas should remain bound to the BH. 
Numerous observations indicate the presence of $10^4 \mathrm{M_{\odot}} - 10^5 \mathrm{M_{\odot}}$ IMBHs in galactic nuclei. 
One of the most convincing observations of this type reveals
that the nuclear star cluster of the galaxy NGC 4395  hosts a BH with a mass between $9\times10^3 \mathrm{M_{\odot}}$ \citep{Woo2019} up to $\sim 4 \times 10^5 \mathrm{M_{\odot}}$ \citep{Peterson2005, denBrok2015}. 
Our results show that tidal captures and tidal disruption events might contribute significantly to the growth of the many observed IMBHs inside nuclear star clusters.
TCEs and TDEs might even be the most dominant IMBH growth channel in clusters with moderate escape velocities ($<100$ km/s), where the hierarchical formation scenario is suppressed by relativistic recoils \citep{Mapelli2021}.  \\
In this work, we did not intend to conduct realistic simulations of NSCs. Instead, we developed idealised systems that contain all the key ingredients to address the questions we aim to answer while being an approximate version of real systems. There are two main simplifications adopted in this study.
First of all, we initialised our systems with an old population of stars. In this way, we were able to neglect effects related to stellar evolution and focus exclusively on dynamic effects. However,  NSCs experience multiple episodes of star formation that replenish the system of massive stars \citep{Carson2015, Kacharov2018}. The latter might contribute significantly to the BH mass growth and could dramatically alter the scenario described in this work.
In addition, our simulations contain only a single central BH. In other words, we assume that after the BH subsystem evaporates, only one BH remain in the cluster.
However, the stellar system might keep a BH binary instead of a single BH which might significantly change the tidal disruption rate experienced by the compact objects. \\
In future studies, we plan to include in our initial conditions the central BHs subcluster, since the latter could significantly influence the IMBH tidal disruption rate \citep{Teboul2022}.
We also plan to locate our systems at the centre of an external potential to study the effect of the galaxy bulge around a NSC.
In addition, in subsequent work, we intend to make a comprehensive comparison between our findings with analytical models for loss cone dynamics \citep{Stone2016} and tidally-driven runaway growth \citet{Stone2017}.

\section*{Acknowledgements}
FPR, PHJ, SL and DI acknowledge the support by the European Research Council via ERC Consolidator Grant KETJU (no. 818930). PHJ also acknowledges the support of the Academy of Finland grant 339127. 
TN acknowledges support from the Deutsche Forschungsgemeinschaft (DFG, German Research
Foundation) under Germany's Excellence Strategy – EXC-2094 –
390783311 from the DFG Cluster of Excellence 'ORIGINS'.
The simulations have been carried out with the Ampere GPU system of the MPI for Astrophysics cluster "Freya" hosted by the Max Planck Computing and Data facility. This study used also facilities hosted by the CSC - IT Centre for Science, Finland.

\appendix

\section{The properties of the bound cloud}\label{apx:bound_cloud}

\begin{figure*}
    \includegraphics[width=0.98\textwidth]{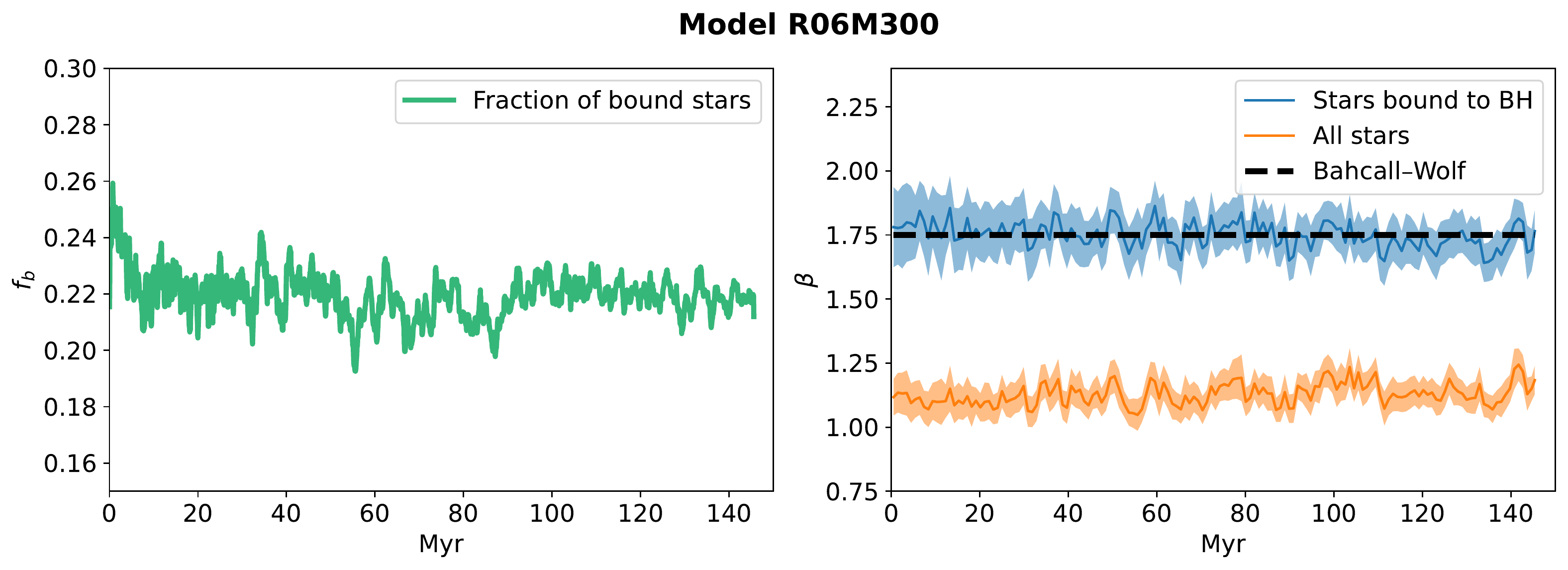}
    \caption{\textit{Left panel}: The same description as in Fig. \ref{fig:7_binary_fraction}. The fraction of bound stars stays constant for the entire evolution of the model R06M300. \textit{Right panel}: The same description as in Fig. \ref{fig:5_BW_comparison}. The bound cloud density profile in R06M300 is consistent with a BW density profile throughout the simulation.}
    \label{fig:Appendix_R06M300}
\end{figure*}

\begin{figure*}
    \includegraphics[width=0.98\textwidth]{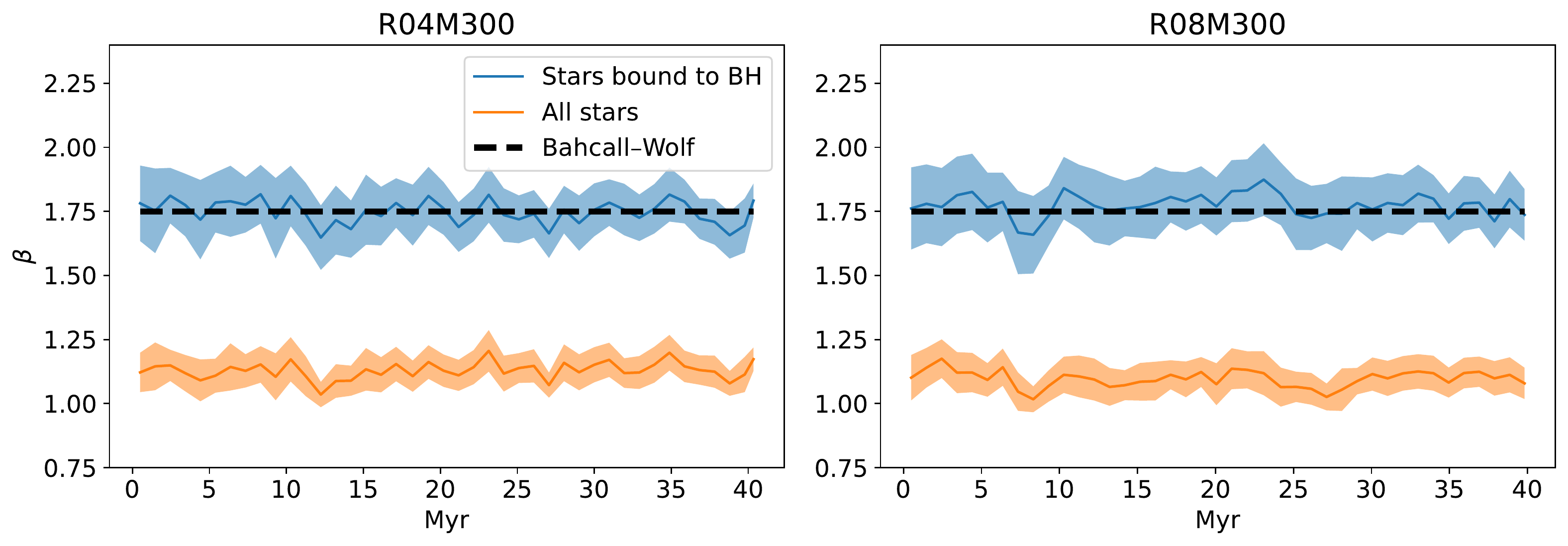}
    \caption{The same description as in Fig. \ref{fig:5_BW_comparison}. Here we compare the value of the slope parameter $\beta$ as a function of time for two models with different initial half mass radii: R04M300 (left panel) and R08M300 (right panel). The two models appear to have almost identical behaviour, and in both cases, the bound subsystem density profile is consistent with the BW distribution.  
    }
    \label{fig:Appendix_BW}
\end{figure*}

In this paper, we show that the central BHs form a cloud of bound stars within their influence radius ($r<\Rinf$). Apart from an initial phase, the fraction of bound stars tends to a constant value in all simulations, as we see Fig. \ref{fig:7_binary_fraction}. For simplicity, in this figure, we show the evolution of the fraction of the bound stars $f_{\mathrm{b}}$ only until $40$ Myr.   Since we evolve the model R06M300 for $\Tend$ here (see left panel of Fig. \ref{fig:Appendix_R06M300}), we report that $f_{\mathrm{b}}$ stay roughly constant also for longer time periods. Moreover, in the main text, we indicate that stars in the bound cloud assume a spatial distribution compatible with a Bahcall-Wolf density profile. Also, in this case, for the sake of simplicity and clarity, we show how the Bahcall-Wolf develops only for the models  R06M50, R06M300 and R06M2000 for $t \leq 40$ Myr. In this appendix, we reveal that the cusp persists for a time scale beyond $40$ Myr as displayed in the right panel of Fig. \ref{fig:Appendix_R06M300}. Moreover, we show in Fig. \ref{fig:Appendix_BW} that also models with different initial concentrations develop a Bahcall-Wolf cusp.

\section{Fitting formulas for the cluster core evolution}\label{apx:extrapolations}

\begin{figure*}
    \includegraphics[width=0.98\textwidth]{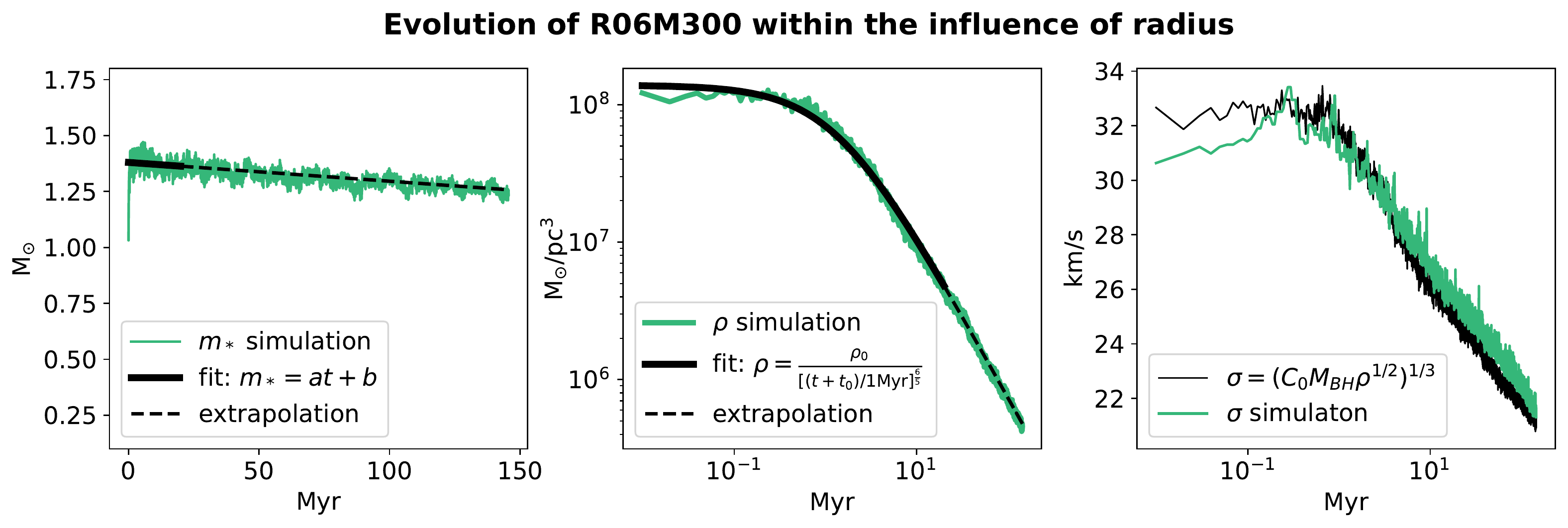}
    \caption{\textit{Left panel}: Average mass of bound particles enclosed within the influence radius as a function of time (green line). Best linear fit with parameters $a, b$ calibrated using the first 20 Myr of evolution (continuous black line). Extrapolation of the linear fit until the end of the simulation (black dashed line). \textit{Central panel}: R06M300 density at $r=\Rinf$ as a function of time (green line). Best fit using Eq. \ref{eq:rhofit} with parameters calibrated using the first 20 Myr of the evolution (continuous black line). Extrapolation of the fit (black dashed line).  \textit{Right panel}: R06M300 velocity dispersion within the influence radius as a function of time (green line). Predicted velocity dispersion using equation \ref{eq:sigmafit}. }
    \label{fig:Appendix_extrapolations}
\end{figure*}

In subsection \ref{sub:mass_growth}, we describe how two-body relaxation processes drive bound orbits to tidal disruption, and we estimate the rate of TDEs with $\NdotTDE  \propto \frac{\Nb}{\trel}$.
Using this relation, we derived the TDE rate as a function of $m_*$, $\rho$ and $\sigma$ calculated at the influence radius (see Eq. \ref{eq:Ndot_2}).
In this appendix, we predict the time evolution of these quantities for the model R06M300 employing fitting formulas, which we then use to extrapolate the mass growth of the BH for about 1 Gyr. 

The green line in the right panel of Fig. \ref{fig:Appendix_extrapolations} indicates that $m_*$ decrease linearly over time. We therefore use the fitting formula $m_*(t)= a+bt$.
 We calibrate the parameters $a=1.38 \mathrm{M_{\odot}}$ and $b=-8.4\times10^{-4}\frac{\mathrm{M_{\odot}}}{\mathrm{Myr}}$  using the first 20 Myr of the simulation
 R06M300. The black line in Fig. \ref{fig:Appendix_extrapolations} (right panel) shows the fit, while the dashed black line shows the extrapolation of the fit, which is also in good agreement with the simulation.
 From empirical considerations we realised that the evolution of $\rho$ as a function of time, can be modelled using the following fitting function:
\begin{equation} \label{eq:rhofit} 
\rho(t)= \frac{\rho_0}{[(t_0+t)/1\, \mathrm{Myr} ]^{\frac{6}{5}}}
\end{equation} 
The best parameters to fit the central density evolution of R06M300 are  $\rho_0=1.94\times10^8 \frac{\mathrm{M_{\odot}}}{\mathrm{pc^3}}$ and $t_0=1.3$ Myr. 
Also, in this case, the parameters are found using the first 20 Myr of the simulation.
The central panel in Fig. \ref{fig:Appendix_extrapolations} indicates that equation \ref{eq:rhofit} does an excellent job of extrapolating the simulation density for a longer time. 
Regarding $\sigma$, we derive using the virial theorem that:
\begin{equation} \label{eq:sigmafit}
\sigma = \left(C_0 \left(\frac{\Mbh}{10^3 \mathrm{M_{\odot}}}\right) \sqrt{\frac{\rho}{10^7\mathrm{M_{\odot}}/\mathrm{pc}^3} }\right)^{\frac{1}{3}} \mathrm{km/s}
\end{equation}
where $C_0$ is a dimensionless constant that depends on the distribution of the stars within the influence radius. For the model R06M300 we find $C_0=3.33\times10^{4}$.

\bibliographystyle{mnras}
\bibliography{output}

\bsp	
\label{lastpage}

\end{document}